\documentclass[12pt,preprint]{aastex}

\newcommand{\Msolar} {M_\odot}

\newcommand{\Teff} {T_{\rm eff}}
\newcommand{\logg}{\log g}
\newcommand{\logeps}{\log \epsilon}
\newcommand{\vsini}{v \sin i}
\newcommand{\vrot}{v_{\rm rot}}
\newcommand{\kms}{\; {\rm km} \; {\rm s}^{-1}}
\newcommand{\K}{\; {\rm K}}
\newcommand{\unit}[1]{\,\,{\rm #1}}
\newcommand{\I}{~{\scshape i}}
\newcommand{\II}{~{\scshape ii}}
\newcommand{\III}{~{\scshape iii}}
\newcommand{\loggf}{\log g\!f}

\newcommand{\mi}{$-$}
\newcommand{\pl}{$+$}
\newcommand{\lt}{$<$}
\newcommand{\fmi}{\phantom{$-$}}
\newcommand{\fn}{\phantom{0}}
\newcommand{\e}{$\pm$ } 

\newcommand{\etal}{{\it et al.}}
\newcommand{\MV}{$M_V$}
\newcommand{\BV}{$B\!\!-\!\!V$}
\newcommand{\UB}{$U\!\!-\!\!B$}
\newcommand{\UV}{$U\!\!-\!\!V$}
\newcommand{\VI}{$V\!\!-\!\!I$}
\newcommand{\VK}{$V\!\!-\!\!K$}

\shortauthors{Behr}
\shorttitle{Abundances and rotations of BHB stars}

\begin{document}

\title{Chemical Abundances and Rotation Velocities of Blue Horizontal-Branch Stars in Six Globular Clusters \altaffilmark{1}}
\author{Bradford B. Behr \altaffilmark{2}}
\affil{Palomar Observatory \\ Mail Stop 105-24, California Institute of Technology \\ Pasadena, CA, 91125}

\altaffiltext{1}{Based on observations obtained at the W.M. Keck Observatory, which is operated jointly by the California Institute of Technology and the University of California.}
\altaffiltext{2}{Present address: McDonald Observatory, RLM 15.308, University of Texas at Austin, Austin TX 78712,  email {\tt bbb@astro.as.utexas.edu}}

%%%%%%%%%%%%%%%%%%%%%%%%%%%%%%%%%%%%%%%%

\begin{abstract}

High-resolution spectroscopic measurements of blue horizontal-branch stars in six metal-poor globular clusters --- M3, M13, M15, M68, M92, and NGC~288 --- reveal remarkable variations in photospheric composition and rotation velocity as a function of a star's position along the horizontal branch. For the cooler stars ($\Teff < 11200 \K$), the derived abundances are in good agreement with the canonical cluster metallicities, and we find a wide range of $\vsini$ rotation velocities, some as high as $40 \kms$. In the hotter stars, however, most metal species are strongly enhanced, by as much as 3 dex, relative to the expected cluster metallicity, while helium is depleted by 2 dex or more. In addition, the hot stars all rotate slowly, with $\vsini < 8 \kms$. The anomalous abundances appear to be due to atomic diffusion mechanisms --- gravitational settling of helium, and radiative levitation of metals --- in the non-convective atmospheres of these hot stars. We discuss the influence of these photospheric metal enhancements on the stars' photometric properties, and explore possible explanations for the observed distribution of rotation velocities.

\end{abstract}


%%%%%%%%%%%%%%%%%%%%%%%%%%%%%%%%%%%%%%%%

\section{Introduction}

The chemically homogeneous and coeval stellar populations found in Galactic globular clusters (GCs) present many excellent opportunities for testing our understanding of stellar evolution. Observations of the photometric properties of stars at different evolutionary stages, represented by the various sequences and branches that appear in a cluster's color-magnitude diagram (CMD), can be compared to theoretically predicted isochrones and loci, not only to evaluate the accuracy and completeness of the models, but also to derive fundamental characteristics of the clusters themselves, such as distance and age. Stars on the horizontal branch (HB) are especially useful for probing a variety of aspects of post-main-sequence evolution because their photometric properties are very sensitive to composition and structure. A particular star's position along the color axis of the HB depends strongly on its metallicity and the amount of mass loss it encountered earlier in its evolution, so the distribution of stars along a cluster's HB locus offers important clues regarding the mechanisms which drive mass loss. In addition, stars that reside on the blue horizontal branch (BHB) and extended horizontal branch (EHB) have lost a significant fraction of their envelopes, and mass layers which were previously deep within the star are now visible at the surface, so we may gain some insights into the internal structure and dynamics of the star at earlier times.

Differences in HB color morphology from cluster to cluster can be partially attributed to differing metallicities, as first noted by \cite{sandage60}. Metal-rich clusters tend to have short red HBs, while metal-poor clusters exhibit predominantly blue HBs. Cluster metallicity, therefore, is considered to be the ``first parameter'' influencing HB morphology. However, some other parameter(s) in addition to metallicity must also be at work, as clusters with nearly identical [Fe/H] can show very different HB color distributions \citep{vandenbergh67, sandage67, rich97, laget98}. This ``second parameter'' was initially thought to be cluster age \citep{searle78, lee94, stetson96}, which determines the cluster turnoff mass and thus sets the subsequent maximum hydrogen envelope mass, but several alternative or additional second-parameter candidates have subsequently been suggested, including helium abundance and mixing \citep{sweigart97}, CNO abundance \citep{rood81}, central concentration of the cluster \citep{fusipecci93, buonanno97}, and distribution of stellar rotation rates \citep{peterson95}. The potential influence of each of these factors is still very much an open question. For a more comprehensive review of the second-parameter problem, see \cite{fusipecci98}.

Detailed observations of the BHBs of metal-poor clusters reveal additional peculiarities. Even among clusters with predominantly blue HB morphologies, the bluewards extent of the BHB can vary significantly; some clusters exhibit long ``blue tails,'' populated by hot BHB and EHB stars, while other clusters have only the cooler BHB stars. Furthermore, high-precision photometry reveals ``gaps'' in the distribution of stars along these blue tails \citep{ferraro97a, ferraro98, sosin97, catelan98, piotto99}, some of them quite wide, some of them narrow and sharply defined. Although a few of these gaps may merely be statistical fluctuations \citep{catelan98}, others could be real features, as they often show up at similar locations in different clusters. Of particular interest is the gap at $\Teff \simeq 11000$--$12000 \K$, labeled ``G1'' by \cite{ferraro98}, which appears to coincide with a ``jump'' in Str\"omgren $u$-band luminosity first discovered in M13 by \cite{grundahl98}. For stars cooler than $\sim 11500 \K$, Grundahl {\it et al.} found good agreement between the observed and theoretical BHB loci, but the hotter stars' $u$ magnitudes were brighter than predictions by $\sim 0.4$~mag. These $u$-jumps have now been found in every one of 14 globular clusters studied by \cite{grundahl99}, suggesting that some mechanism, not included in the stellar models, affects all HB stars hotter than $11500 \K$. Similarly, surface gravities derived from Balmer and helium line profile fitting \citep{moehler95, moehler97a, moehler97b, moehler03} exhibit a systematic offset towards lower $\logg$ for $12000 \K < \Teff < 20000 \K$, while stars outside this temperature range agree well with theoretical predictions. The magnitude of this ``low-gravity jump'' is small, but it appears consistently in a number of different clusters, and has proven difficult to account for via luminosity evolution or higher helium content, again suggesting that the influence of some additional stellar characteristic is not fully accounted for in the models. \cite{moehler01} and \cite{piotto03} provide an excellent reviews of these issues and other related topics.

Over the past 20 years, several different research groups have undertaken detailed spectroscopic studies of individual BHB stars in these metal-poor GCs, providing an opportunity to look for specific differences in stellar characteristics which might explain the observed photometric features. The pioneering work on BHB star rotation was done by \cite{peterson83a, peterson83b, peterson85a, peterson85b, peterson95}, who determined $\vsini$ rotation velocities for cooler (8000--$11000 \K$) BHB stars in six globular clusters plus the metal-poor field population. The most notable aspect of their findings was the difference in maximum observed rotation velocity among the clusters. In M13, some stars were found to be rotating as fast as $40 \kms$, considerably faster than one would expect for the progeny of slowly-rotating late main sequence stars. However, the distribution of $\vsini$ values observed in this cluster strongly suggested a bimodal distribution of true rotation speeds, with roughly one-third of the stars spinning fast (35--$40 \kms$), while the other two-thirds rotate at a slower rate (15--$20 \kms$). A similar fraction of faster-rotating stars was found among the field BHB population, while the other clusters (M3, M4, M5, NGC~288) appeared to possess only the slowly rotating stars. Two fast-rotating BHB stars were also found in M92 by \cite{cohen97}, out of five stars observed. \cite{behr99} then extended the observations of M13's BHB to higher temperatures, and discovered that the fast-rotating population is not present above $11000 \K$ --- the hotter stars in this cluster are all spinning at $\vsini < 8 \kms$. A similar pattern was subsequently reported in M15 by \cite{behr00b}. A more comprehensive survey of bright field BHB stars by \cite{kinman00} turned up additional fast rotators, although a majority of the field stars are of the slowly rotating ($\sim 20 \kms$) variety. Most recently, \cite{recioblanco02} describe the first results from a VLT program to measure BHB $\vsini$ in several southern GCs. In M79, they find a similar pattern as in M13 and M15 --- a substantial fast-rotating population between 8000 and $11500 \K$, but only slow rotation above $11500 \K$ --- while M80 and NGC~2808 show no fast rotators among the cooler stars.

The first hint that photospheric chemical abundances might also change significantly as a function of position on the HB came from \cite{glaspey89}, who measured iron and helium line strengths for two stars in NGC~6752. The cooler one, at $\Teff \simeq 10000 \K$, showed [Fe/H]$= -1.5$, as expected for this cluster, but the hotter star, at $16000 \K$, was found to be strongly enhanced in iron and depleted in helium. More comprehensive surveys of abundances of BHB stars in this temperature range were done by \cite{behr99} in M13, and \cite{moehler99} in NGC~6752. In both cases, a sharp transition from normal abundances to strongly metal-enhanced, helium-depleted photospheres was found at $\Teff \simeq 11500 \K$. Shortly thereafter, \cite{behr00b} found a nearly identical pattern in M15.

In this paper, we undertake a full reanalysis of our complete data set, parts of which were previously published by \cite{cohen97}, \cite{behr99}, \cite{behr00a}, and \cite{behr00b}. In Section 2, we discuss the target sample in more detail, and provide an overview of the observations and spectral reduction procedure. Sections 3 and 4 describe how the stellar photospheric parameters --- $\Teff$, $\logg$, microturbulence $\xi$, $\vsini$, and chemical abundances $\log\epsilon$ --- are determined from the photometric and spectroscopic data. We present our results in Sections 5, 6, and 7, and discuss possible explanations for the observed trends in Sections 8, 9, and 10. Section 11 summarizes the results of this work, and offers some suggestions for future research directions.

%%%%%%%%%%%%%%%%%%%%%%%%%%%%%%%%%%%%%%%%

\section{Target selection, observations, and echelle reduction}

For this project, we observed 74 BHB stars in six metal-poor GCs. Table~\ref{cluster-parameters} summarizes the characteristics of the program clusters, with data drawn from the most recent version of the \cite{harris96} globular cluster catalog. Clusters were selected in order to sample a range of both metallicity and HB morphology --- some long blue tails, some shorter tails, some GCs with hardly any tail at all --- given the limitations of finite observing time.

\begin{figure}
\epsscale{0.85}
\plotone{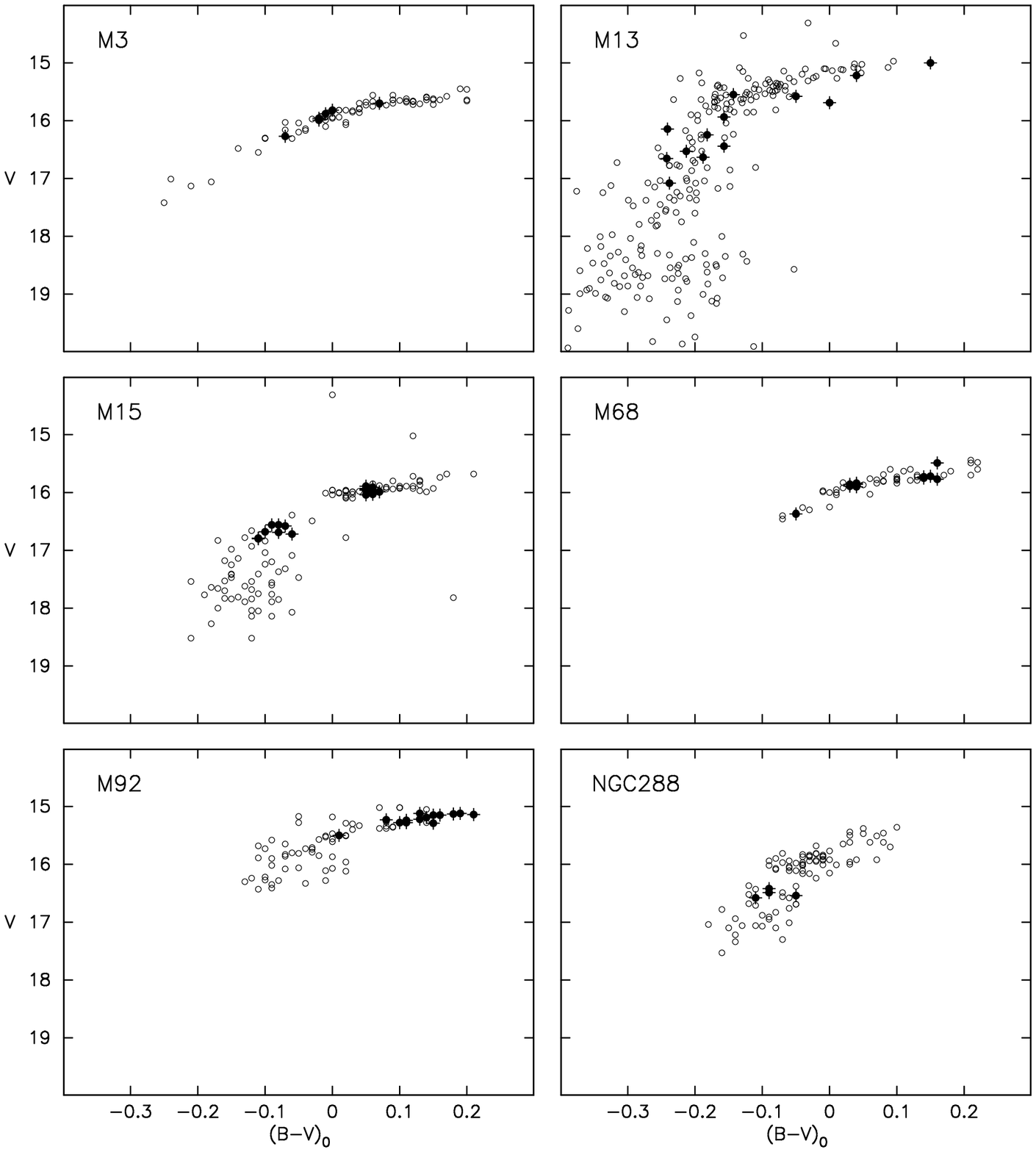}
\caption{CMDs of the HBs of our six globular clusters. Solid symbols with crosshairs denote the target stars for this study. \label{target-cmds}}
\end{figure}

\begin{figure}
\epsscale{0.70}
\plotone{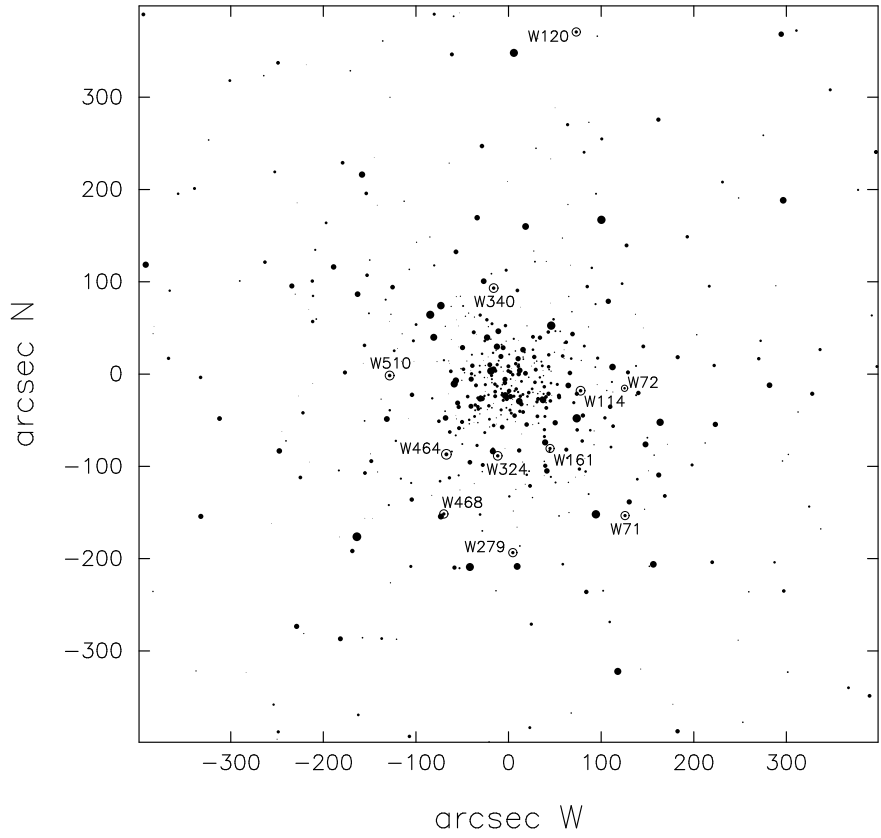}
\caption{Finding chart for M68 targets. North is up, east is left. \label{m68-fc}}
\end{figure}

\begin{figure}
\epsscale{0.70}
\plotone{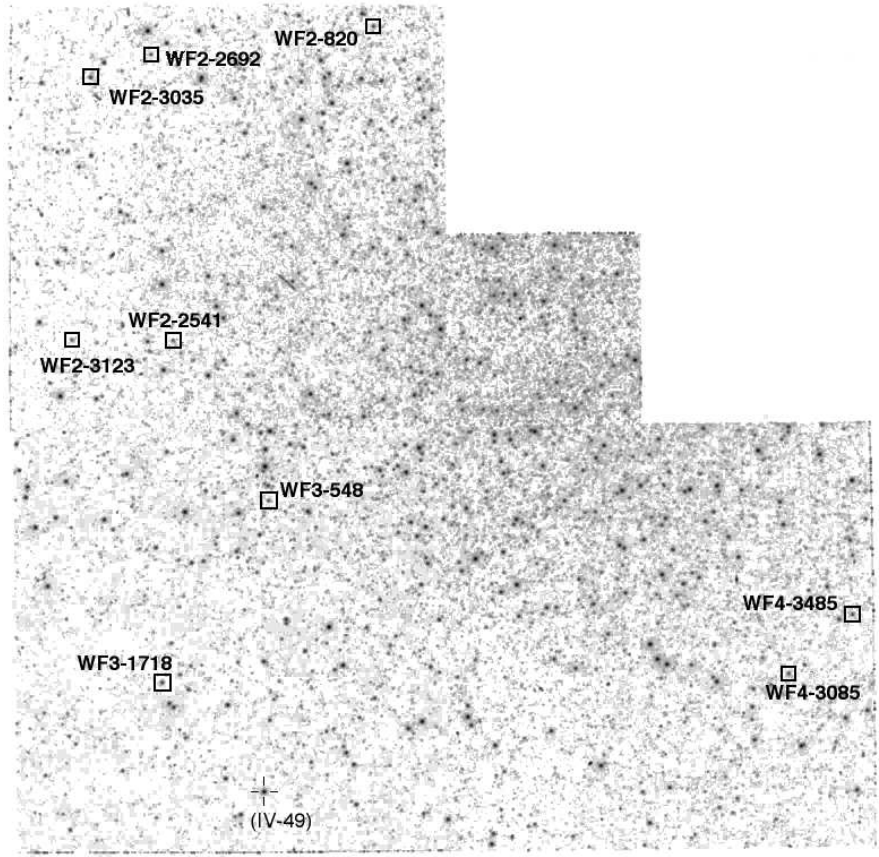}
\caption{Finding chart for M13/WF targets. North is up, east is left. Star IV-49 from \cite{arp55a} is marked as a reference point. \label{m13-wfpc}}
\end{figure}

Within each cluster, we attempted to observe stars spanning a range of stellar effective temperature. In M13 and NGC~288, we concentrated on the hotter end of the BHB, since the cool end of the BHB had been thoroughly covered by \cite{peterson95}. Four stars in M13 and two in M3 are in common with the Peterson target lists, and two of Recio-Blanco's stars in M15 were selected to overlap with our sample, to permit comparison of the rotation results. We show CMDs of the BHB region of each of these six clusters in Figure~\ref{target-cmds}, with our target stars marked. Finding charts for the target sample are available from a variety of sources, summarized in Table~\ref{finding-charts}, and Figures~\ref{m68-fc} and \ref{m13-wfpc} label the target stars in M68 and the core of M13.

All of the spectroscopic data for this project were acquired with the Keck I telescope and its HIRES spectrograph \citep{vogt94}. HIRES is a cross-dispersed echelle instrument, offering high spectral resolution over a large wavelength range. Most of these BHB observations used the C1 slit decker, which measures 0.86~arcsec in the dispersion direction, yielding a 3-pixel spectral resolution element with $R = \lambda/\delta\lambda = 45000$, equivalent to a velocity resolution of $v = 6.7 \kms$. A few of the early observing runs utilized the wider C5 decker, 1.15~arcsec wide, which yields $R=34000, v= 8.9 \kms$. Two different cross-disperser gratings (`RED' and `UV') are available for HIRES, depending on the slit length and order spacing for the spectral region being observed. Since many of the BHB observations were piggybacked on other observing programs, different cross-disperser and echelle grating settings were used from run to run, and spectral coverage varies slightly as a result. Table~\ref{hires-config} provides an overview of the seven different grating configurations used in this program.

In order to get the high signal-to-noise ($S/N$) ratios necessary for the abundance and rotation measurements, we integrated on each star for a total of 1.0 to 1.7 hours. We usually limited individual exposure times to 1200 or 1500 seconds, to minimize cosmic ray accumulation, and then coadded three to four frames per star. The chip readout was binned by two pixels in the spatial direction, to decrease readout time and read noise. In Table~\ref{hires-obs}, we summarize the observations on each target star, including the mean $S/N$ per pixel, as estimated from the rms dispersion of the continuum regions of the final reduced spectrum. At the beginning and end of each night, we also took multiple bias frames and flatfield and arc lamp exposures for the subsequent reduction and calibration procedures.

\begin{figure}
\epsscale{0.80}
\plotone{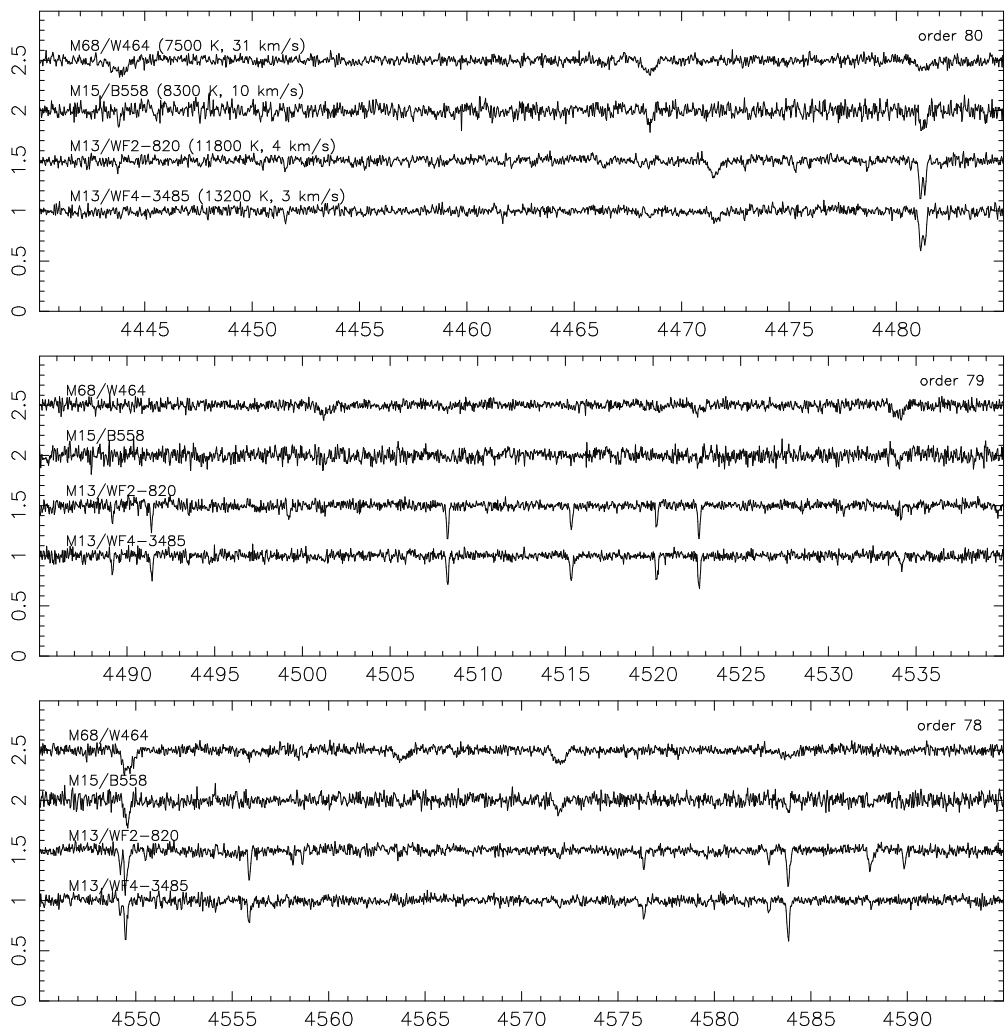}
\caption{Selected orders from the normalized spectra of four target stars, shifted to zero radial velocity, and offset vertically by multiples of 0.5 for clarity. The cooler stars show broader line profiles, due to higher projected rotation velocities. The Mg\II\ 4481 line and assorted Fe and Ti lines are visible in all spectra, and the hotter two stars also show He\I\ 4471, P\II\ 4588, and P\II\ 4590 lines. \label{spectrum}}
\end{figure}

We used a suite of routines \citep{mccarthy90} developed for the FIGARO data analysis package \citep{shortridge93} to reduce the HIRES echellograms to arrays of one-dimensional spectra. CCD frames were first bias-subtracted, using the overscan region from each frame and zero-second dark frames. A master normalized flatfield frame was constructed from several different exposures of the spectrograph's internal incandescent lamps, median-filtered to remove cosmic ray hits, and appropriately weighted to provide uniform response from order to order. Cosmic ray hits on the data frames were identified and removed by hand in order to minimize the potential distortion of narrow spectral features. Each spectral order was traced by a 10th-order polynomial, and the spectrum extracted via direct pixel summation. For the blue wavelength ranges covered, the sky background and scattered light background proved negligible, so no background subtraction was necessary. HIRES uses an internal thorium-argon arc lamp for wavelength calibration, and residuals on the polynomial fit for the wavelength solution for each order averaged 5~m\AA\ or less. The stellar continuum for each order was easily identified for these warm, metal-poor stars, and 6th- to 8th-order polynomial fits provided normalization to unity. Several representative orders from one of our stars are shown in Figure~\ref{spectrum} to illustrate the appearance of the final spectra.

To properly compare synthetic and observed spectra in subsequent spectral analyses, we need to accurately determine the instrumental broadening function of each spectrograph configuration. From each night's calibration data, we identify all the unsaturated thorium-argon emission lines, normalize their amplitudes, stack their line profiles atop one another, and then iteratively discard the profile that differs most from the mean profile, until the rms dispersion of all normalized line profiles is less than 0.01. This median-filtering procedure eliminates blended and otherwise atypical line profiles, and the resulting mean arc line profile gives us a reasonable estimate of the true instrumental profile. As a cross-check, we also perform Gaussian fits to the night-sky emission lines that appear in some of our spectra, and in Table~\ref{nse-cmp}, we compare the resolving power $R$ estimated from the FWHM of those lines to the $R$ deduced from the median thorium-argon line profiles. A majority of the 15 selected night sky lines have widths that closely match those of the arc lamp lines, although three of the lines appear to be significantly narrower, and one is noticeably wider. Nonetheless, it appears that the instrumental profile derived from calibration data closely matches the instrument's spectral resolving power on the sky --- assuming that the slit is uniformly illuminated by the light source. When the seeing is particularly good, the stellar image may ``underfill'' the slit, so that the effective slit width is narrower, and the instrumental profiles derived from arc lamp spectra would overestimate the actual width of the instrumental profile. To evaluate the magnitude of this effect, we de-convolve the empirically-determined instrumental profiles assuming a rectangular slit profile, then re-convolve them with a truncated Gaussian seeing profile, and measure the difference in FWHM between the original and seeing-corrected instrumental profiles. Table~\ref{seeing-test} shows the results of this exercise for both narrow and wide slit deckers, over a range of seeing FWHM. Given the seeing conditions during our observing runs (0.7--1.2~arcsec FWHM for narrow slit, 0.9--1.1~arcsec FWHM for wide slit), we estimate that the underfilling of the slit could account for at most a 10--14\% underestimate of the width of the instrumental profile. We will consider the potential effects of such an underestimation in a subsequent section.

%%%%%%%%%%%%%%%%%%%%%%%%%%%%%%%%%%%%%%%%
			
\section{Photometric analysis}

Photometric data for these six GCs were initially used for selecting BHB targets, but then also played an important role in establishing the photospheric parameters adopted for the abundance analysis. We relied on a variety of different photometric studies from the literature, summarized in Table~\ref{photometry}. To transform from color and magnitude to temperature and gravity, we employed grids of synthetic colors computed from ATLAS9 model atmospheres by R.~Kurucz and made available from his website (\url{http://kurucz.harvard.edu/}). By comparing the observed photometry of each star, appropriately corrected for the extinction and reddening for each cluster, to the predicted color over a grid of atmospheric parameters ($\Teff$, $\logg$, [Fe/H]), we constrain each star's surface temperature and gravity. The photometric colors are most sensitive to temperature, with only a modest dependence on gravity, while the absolute magnitude $M_V$ (computed from the observed $V$ magnitude, with extinction and distance modulus from Table~\ref{cluster-parameters}) provides a more stringent constraint on $\logg$. More specifically, we employ the $AB$ magnitude introduced by \cite{oke83}, which lets us compute a synthetic $M_V$ given the photospheric Eddington flux $H_\nu$ at 5480~\mbox{\AA} from a computed model atmosphere. We start with Oke \& Gunn's Equation~1 (correcting the sign of the final constant, which was misprinted in the original paper):
	\[ V \simeq AB = -2.5 \log f_\nu(5480\,\mbox{\rm\AA}) - 48.60 \]
For a star at a distance of 10~pc, $M_V = V$, and the flux at the telescope $f_\nu$ is equal to the flux at the stellar surface $F_\nu = 4\pi H_\nu$, scaled by the ratio of stellar radius $R_*$ to stellar distance:
	\[ f_\nu = F_\nu \left( {R_* \over 10{\rm ~pc}} \right)^2 = H_\nu R_*^2 \times {4 \pi \over (10{\rm ~pc})^2} \]
The radius of the star can be cast in terms of stellar mass $M_*$ and surface gravity $g$:
	\[ R_*^2 = {G \Msolar \over g} \times {M_* \over \Msolar}  \]
and substituting into the first equation, we get:
	\[ M_V = -2.5 \left( \log H_\nu + \log(M_*/\Msolar) - \log g + \log {4\pi G\Msolar \over (10{\rm ~pc})^2} \right) - 48.60 \]
which simplifies to:
	\[ M_V = 2.5 \left( \log g - \log H_\nu - \log (M_*/\Msolar) - 7.68 \right) . \]
Using this last equation, we can calculate a grid of $M_V$ over a range of temperature, gravity, metallicity, and stellar mass, drawing the value of $H_\nu(5480\,\mbox{\rm\AA})$ from the model atmosphere for the specified $\Teff$, $\logg$, and [Fe/H]. We initially assume a typical BHB mass $M_* = 0.6 M_\odot$ and [Fe/H] from the published cluster metallicities. For each subsequent iteration, we then estimate a new $M_*$ by comparing the derived temperature and cluster metallicity to the HB models of \cite{lee94}, and use [Fe/H] from the abundance analysis to select the appropriate metallicity for the synthetic photometry grids. We estimate that uncertainties in temperature and metallicity could influence the adopted mass for a star by as much as $0.06 \Msolar$, or $\Delta\log(M_*/\Msolar) = 0.04$, yielding an equivalent change in $\logg$ of 0.04~dex, which would constitute only a small contribution to the total error budget for $\logg$.

\begin{figure}
\epsscale{0.70}
\plotone{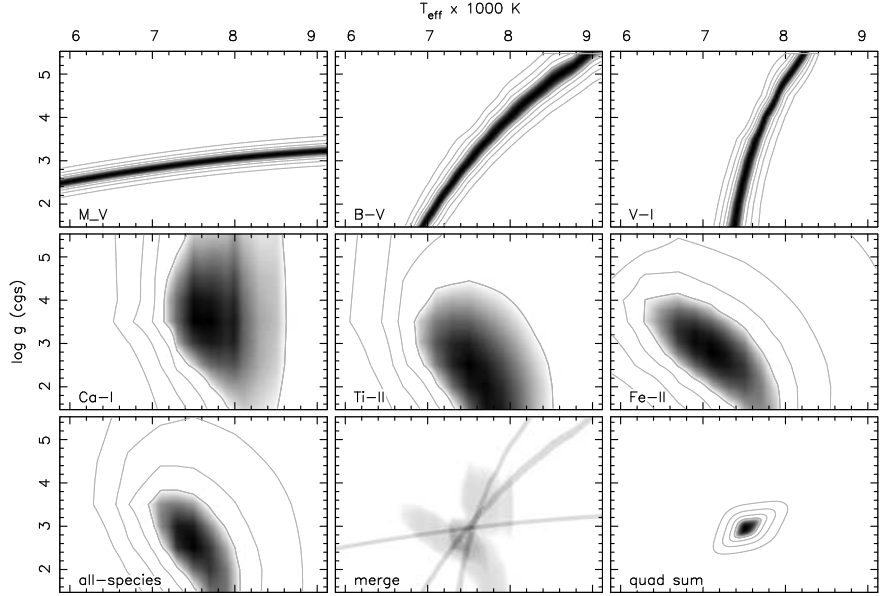}
\caption{$(\Teff, \logg)$ solution for cool star M68/W464. The $z$-functions for each photometric and spectroscopic constraint define different solution swaths through the parameter plane ($\Teff$ in thousands of K on the $x$-axis, $\logg$ in cgs units on the $y$-axis). Greyscale denotes the range $z = 0$ (darkest) to $1$ lightest, with dark contour lines at $z = 0.25, 1, 2, 4, 8$ and light contour lines at $z = 16, 32, 64, 128$. The top three panels show the grids computed from photometric colors and absolute magnitude, while the middle three panels show the grids resulting from spectroscopic analysis of lines of Ca\I, Ti\II, and Fe\I\  respectively, and the bottom left panel (``all species'') is the quadrature sum of the three spectroscopic grids. In the center bottom panel (``merge''), all the preceding $z$-functions are overplotted to illustrate their relative positions in the $\Teff-\logg$ parameter plane, and the bottom left panel (``sum'') displays the quadrature sum of all six $z$-function maps, defining an error region for a solution. \label{phot-zmap-1}}
\end{figure}

\begin{figure}
\epsscale{0.70}
\plotone{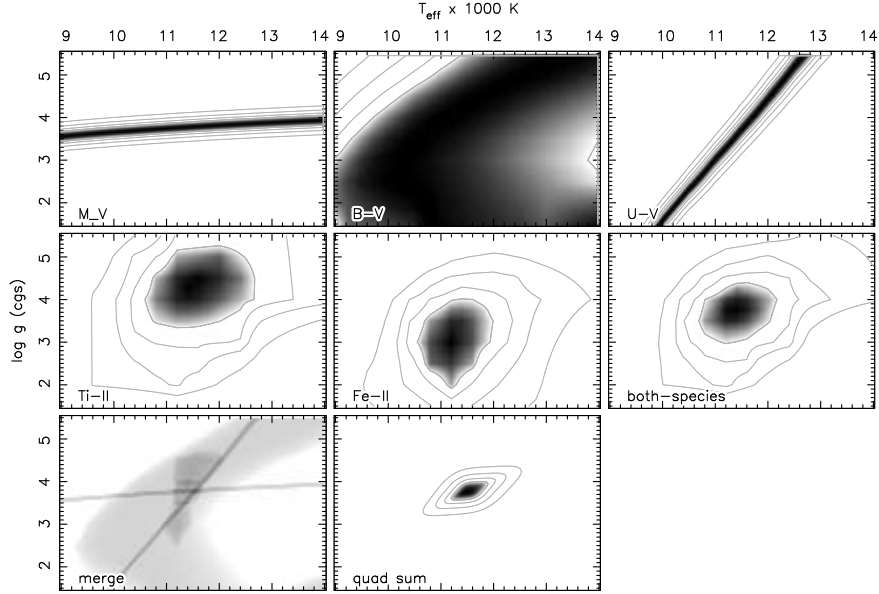}
\caption{$(\Teff, \logg)$ solution for hot star M13/WF3-1718, with a similar representation as the preceding figure. \label{phot-zmap-2}}
\end{figure}

For each observed photometric datum, we compute a map in the $(\Teff, \logg)$ plane, showing how closely the synthetic color or magnitude agrees with the observed quantity. We define a quality-of-agreement parameter $z = (m_{\rm obs} - m_{\rm syn})^2/\sigma_{\rm obs}^2$, such that $z = 0$ at points where the observed and synthetic photometry are in exact agreement, and $z = 1$ at the boundaries of the $1\sigma$ confidence interval, somewhat analogous to a reduced $\chi^2$ statistic. When plotted as a greyscale or contour image, these maps show the locus of possible $(\Teff, \logg)$ solutions permitted by each photometric datum. A composite $z$, computed as the normalized quadrature sum of all the maps, shows the solution point that best satisfies all the constraints, with surrounding confidence contours that delineate the $1\sigma$ error bars on $\Teff$ and $\logg$.  Figures~\ref{phot-zmap-1} and \ref{phot-zmap-2} illustrates this $z$-map procedure for two of our stars.

\begin{figure}
\epsscale{0.80}
\plotone{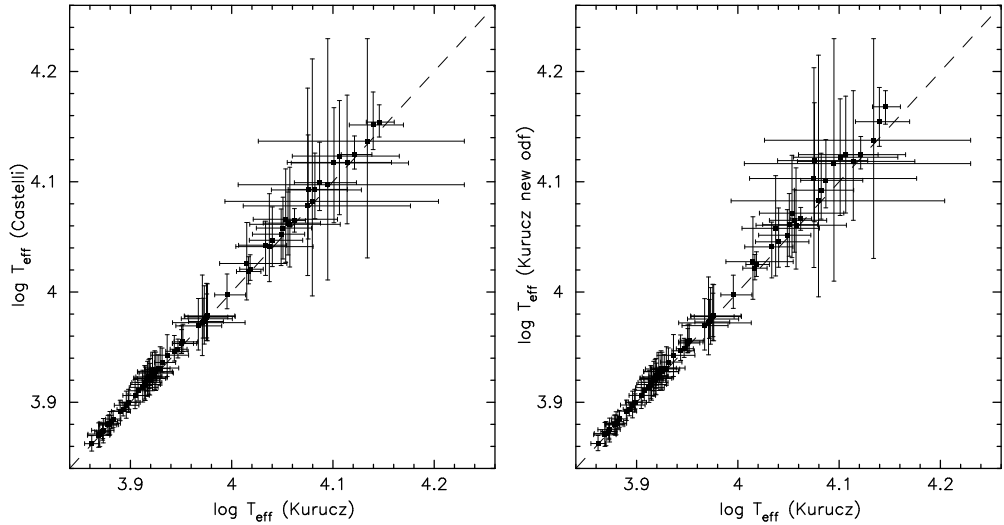}
\caption{Comparisons of $\Teff$ values derived from photometry alone, using different synthetic photometry grids. On the left, we compare the results of the original Kurucz grids to those calculated using the grids of \cite{castelli99}. On the left, the original Kurucz grids are compared to updated Kurucz models with new opacity distribution functions. In both cases, the different grids return very similar results, with only a slight systematic offset towards higher temperatures derived from the newer grids. \label{cmp-phot-src}}
\end{figure}

This $z$-map technique automatically accounts for random errors in the photometric data, but we must also worry about sources of systematic error, such as incorrect cluster distance modulus and reddening, and imperfect synthetic colors. In most cases, these error sources can be incorporated into the uncertainty for the equivalent photometric parameter --- we adopt an error bar of 0.10~mag for $M_V$ for all stars, for instance, to account for errors in the distance modulus and extinction, and quadratically add an estimated error of 0.01~mag in $E(B-V)$ to the photometric measurement error in $B-V$ for each star. This procedure yields useful conservative error intervals for the derived parameters $\Teff$ and $\logg$, but does not truly address the systematic nature of many of these error sources. In order to evaluate variations in the atmosphere modelling, we repeat our photometric analysis with two other sources of synthetic photometry. Figure~\ref{cmp-phot-src} compares the $\Teff$ values derived from the original Kurucz ATLAS9 synthetic photometry grids to the $\Teff$ values computed using photometry grids described by \cite{castelli99}, and Kurucz's ``ODFNEW'' grids. (These newer grids, like the original, are available at \url{http://kurucz.harvard.edu/}). All three data sets use the ATLAS9 code to calculate the spectral energy distribution from a star of given temperature, gravity, and metallicity, but they differ in their treatment of convection and in the opacity distribution functions (ODFs) employed. Use of the newer grids results in $\Teff$ values that are systematically higher than our original results, but the discrepancy is approximately 1\% in $\Teff$, considerably smaller than the random errors, so we believe that the original Kurucz grids are adequate for this analysis.  

To fully address the concerns regarding systematic effects in the photometric analysis, we need a completely independent method for determining the parameters of interest. For many of the program stars, we can compare the photometrically-derived $\Teff$ and $\logg$ for each star to the constraints placed on these parameters by spectroscopic analysis, as described in the next section. We present these comparisons in section 5 below.

%%%%%%%%%%%%%%%%%%%%%%%%%%%%%%%%%%%%%%%%

\section{Spectroscopic analysis}

In order to simultaneously solve for chemical abundances, microturbulent velocity $\xi$, and projected rotation velocity $\vsini$, we have developed an iterative spectral synthesis fitting routine, which scans through a range of parameter space in batch mode, comparing spectra calculated by the LINFOR/LINFIT spectral synthesis package (\cite{baschek66}, subsequently modified by M. Lemke) to the observed spectra. Thermal, turbulent, and rotational broadening are included in the line profiles computed by LINFOR, and the synthetic spectra are then convolved by the instrumental broadening profile computed from the arc lines as described previously, and mapped into the wavelength scale of the observed spectrum. We also make minor adjustments to the continuum normalization, using the synthetic spectra to identify line-free regions in the observed spectrum, interpolating linearly across line features, and then smoothing heavily (3 iterations with a boxcar equal to twice the mean metal line FWHM) to yield a smooth continuum fit.

We use the compilation of \cite{hirata94} as the master spectral line list, and assemble a customized line list for each observed stellar spectrum, by calculating the predicted equivalent width for every transition within the wavelength range of the observations, and discarding those that are too weak (typically $W_\lambda < 5$~m\AA), or which are thought to have poorly-constrained $\loggf$ values (according to the $g\!f$ quality column). For a few lines, the atomic parameters from Hirata \& Horaguchi produced synthetic profiles that were much too strong or too weak compared to other lines of the same species, so we used $\loggf$ values from the VALD database \citep{piskunov95, ryabchikova99, kupka99} instead, which yielded much better agreement. The atomic transitions used for the abundance analysis of each star are listed in Table~\ref{linelist}. Before each iteration, we re-compute this customized line list using the values of $\Teff$, $\logg$, $\xi$, and $\logeps$ derived in the prior iteration.

Model atmospheres are automatically generated as needed by invoking the ATLAS9 program \citep{kurucz93}. The input parameters for ATLAS9 include $\Teff$ and $\logg$, $\xi$ (set to $2 \kms$ initially, and then subsequently updated according to the results of the microturbulence analysis described below), metallicity (based on the [Fe/H] computed from the previous abundance analysis iteration), and helium abundance (also chosen according to the [He/H] computed in the previous iteration). For elements other than iron and helium, we assume solar-scaled abundances; although ATLAS9 is theoretically capable of computing atmospheres with any arbitrary abundance pattern, this is prohibitively difficult in practice. Future analyses of these stars (particularly the hot HB objects with significantly non-solar abundance patterns) might benefit from more detailed atmosphere modelling, if warranted by higher-quality spectral data. We use the default ``overshooting'' setting for ATLAS9's treatment of convection; tests by \cite{castelli97} show that non-overshooting models sometimes yield better agreement between synthetic and observed spectra, but according to their Table~3, the models for our BHB stars should not be strongly affected by the choice of convective treatment.

Chemical abundances for each atomic species are computed by stepping through several different values of $\logeps$ for that species, computing synthetic line profiles for each transition found in the star's line list, and measuring the agreement between the observed and synthetic spectra, as quantified by a spectroscopic quality-of-fit parameter $z$:
\[ z = \sqrt{N_{\rm points}/2} \left( {{\rm rms}^2 \over {\rm rms}^2_{\rm min}} - 1 \right) \]
where $N_{\rm points}$ is the number of spectral data points in the line profile, ``rms'' is the rms deviation between observation and theory computed for those points, and ``rms${}_{\rm min}$'' is the smallest value of rms found. This $z$ parameter is quite similar to a reduced $\chi^2$ measure, but does not require an independent error value $\sigma_i$ for each spectral data point. Plotting $z$ as a function of $\logeps$ usually yields a smooth parabolic curve, with a well-defined minimum point ($z = 0$, where ${\rm rms} = {\rm rms}_{\rm min}$), which indicates the $\logeps$ that best fits all the lines of this species. The $\logeps$ values where $z=1$ (analogous to $\chi^2 = \chi^2_{\rm min} + 1$) define the $1\sigma$ confidence interval in $\logeps$. Although more computationally intensive than calculating $\logeps$ from measured equivalent widths $W_\lambda$, this iterative synthesis approach offers several advantages. Blended lines are handled in the same fashion as unblended lines, and an accurate $\logeps$ for a single blend component can be determined assuming that $\logeps$ for the other species in the blend are well-constrained by other lines, as is usually the case in our uncrowded spectra. Furthermore, we can place stringent upper bounds on $\logeps$ for species that do not appear in the spectrum, by increasing the $\logeps$ value until the synthesized line profiles deviate from the observed flat continuum by a statistically significant amount, $z = 1$.

The radial velocity $v_r$ of each star is initially estimated by manually marking the observed wavelengths of easily-identified strong lines, such as Balmer $\beta$, Balmer $\gamma$, and Mg~II 4481. Once a reasonably accurate synthetic spectrum is available, however, it can be cross-correlated with the observed spectrum to yield a more accurate $v_r$. These telescope-centric velocities are then shifted into a heliocentric frame for comparison to other cluster radial velocity data. On the basis of $v_r$, all of our target stars appear to be members of their associated clusters.

By computing $z$ over a 2-dimensional grid, as a function of both $\logeps$ and $\vsini$, we can find the value of $\vsini$ that provides the best line profile fit for all the lines of a given species. We adopt a value for the limb darkening parameter from Figure 17.6 of \cite{gray92}, using $\lambda = 5000$~\AA\ and the star's $(B-V)_0$ color. The algorithm then steps through a range of $\vsini$ values, and at each point, varies $\logeps$ to find the local minimum in the rms deviation between observed and synthetic line profiles. For the ``correct'' value of $\vsini$, we will find a global minimum rms value, while values of $\vsini$ that are too small or too large will result in poorer line profile fits, and higher values of local minimum rms. By tabulating this local minimum rms as a function of $\vsini$, and then converting into $z$, we find the best-fit value and confidence interval for $\vsini$. Each species is handled separately, so only one $\logeps$ needs to be varied at a time, but then the $z$ vs. $\vsini$ curves for all species are combined (by adding the minimum rms values at each $\vsini$ value) to find the best-fit $\vsini$ value for the spectrum as a whole. This technique permits us to derive good $\vsini$ values from blended lines such as Mg\II\ 4481, which proves useful for metal-poor spectra with few lines.

A similar approach lets us also solve for microturbulence velocity $\xi$, stepping through a grid of $\xi$ values and finding the best-fit rms, with $\logeps$ as a free parameter. Instead of matching the line profile shapes as the $\vsini$ scan does, this scan will find the $\xi$ that gives the best agreement between synthetic and observed line strengths for all the lines of a given species. If $\xi$ is too large or too small, then a single $\logeps$ will not provide a good fit for both the strong and the weak lines of the species, and a higher rms will result. At or near the correct value of $\xi$, all the lines will be well-fit by a single value $\logeps$, and we will find a minimum in the rms vs. $\xi$ curve. As with the $\vsini$ scans, the rms curves for each species are then combined into a master rms and $z$ curves for the star. This rms minimization approach is similar to the traditional technique, in line equivalent width analysis, of choosing the $\xi$ which yields no linear trend in $\logeps$ vs. $W_\lambda$, and the $\xi$ values and error bars that we derive using the synthesis approach agree closely with the values derived from prior equivalent width analysis of the same data. 

We must also adopt a value for the macroturbulent velocity $v_{\rm macro}$, which will affect the best-fit $\vsini$ value for the narrow-lined stars. Measurement of large-scale velocity fields from disk-integrated spectra requires much higher spectral resolution and $S/N$ than our data offer, so we make the assumption that the velocity fields in the stellar photosphere will be of similar magnitude at both smaller and larger length scales, and set $v_{\rm macro} = \xi$ after each $\xi$ scan. This is a gross oversimplification, of course, as the velocity spectrum of turbulent motions in these stellar atmospheres is unlikely to be flat, but this procedure does reflect the expectation that the stable radiative atmospheres of the hotter stars should have $\xi = v_{\rm macro} \simeq 0 \kms$, while $\xi > 0 \kms$ and $v_{\rm macro} > 0$ for the convective envelopes of the cooler stars. In calculating the error budget on derived $\vsini$ (below), we permit $v_{\rm macro}$ to vary by $\pm 1 \kms$, and find that the resulting change in $\vsini$ is small compared to the error range from the spectral fits themselves, so choice of $v_{\rm macro}$ does not appear to be critical.

The $z$-scan technique can be extended yet further, using the different temperature and pressure sensitivities of different lines of a single species to constrain the $\Teff$ and $\logg$ of the star. For each point on a grid in the $(\Teff, \logg)$ plane, we let $\logeps$ vary to find the local best-fit rms. If the species has lines of both high and low excitation potential $\chi$, and the model $\Teff$ and $\logg$ deviate significantly from their true values, then the synthetic spectrum will not agree with the data for all the lines, and a larger rms will result, while at the correct $\Teff$ and $\logg$, the lines will all agree at a single $\logeps$, and we will find a global minimum rms. Converting this grid of rms values into $z$ values, and plotting $z$ as a greyscale or contour map, we see which regions of the $(\Teff, \logg)$ plane provide good solutions for each species, and by adding all the $z$ maps together, a global spectroscopically-derived solution for photospheric temperature and gravity can be determined, providing an independent check on the temperatures and gravities derived from photometry. Figures~\ref{phot-zmap-1} and \ref{phot-zmap-2} show the $(\Teff, \logg)$ solution regions delineated by metal species (middle row of plot panels) for two example stars. We should emphasize that each of these maps is derived from a single atomic species, and does not rely upon assumptions of ionization equilibrium ({\it i.e.} $\logeps($Fe\I$) = \logeps($Fe\II$)$), which are susceptible to non-LTE effects, although we do use ionization equilibrium to test the final photospheric parameters, as described below.

\begin{figure}
\epsscale{0.70}
\plotone{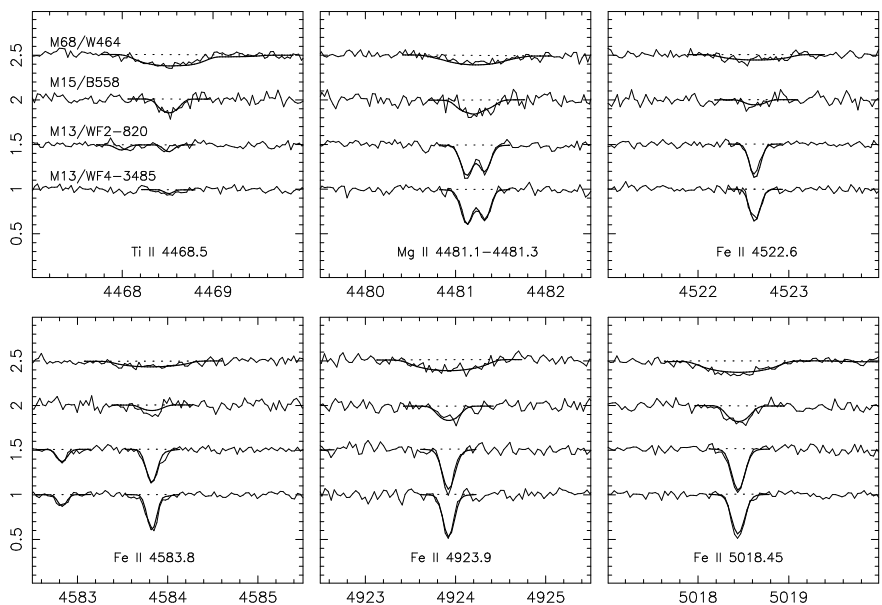}
\caption{Sample line profile fits for six metal lines in four stars, with best-fit synthetic profiles (thick lines) superposed on the normalized observed spectra (shifted to zero Doppler velocity, and offset vertically). \label{synth-fits}}
\end{figure}

Many of these photospheric parameters are highly interdependent, of course --- the best-fit $\vsini$ will depend on the value adopted for $\xi$ (and on the choice of limb darkening parameter and macroturbulent velocity, as discussed below) and vice versa, and $\xi$ and $\logg$ are partially degenerate for many species. In order to find the optimal solution for all parameters, we iterate repeatedly, solving for $\Teff$ and $\logg$, then $\vsini$, then $\xi$, then $\logeps$ for all relevant species. With additional constraints on temperature and gravity from the photometry, as described previously, the results converge within two or three iterations. Figure~\ref{synth-fits} compares the observed and synthetic line profiles for four stars after their final iterations.

We also attempted to use Balmer $\beta$, $\gamma$, and $\delta$ line profiles to constrain $\Teff$ and $\logg$, as is often done with medium-resolution spectra of hot stars. The observed profiles rarely fit any synthetic profiles generated by either BALMER \citep{kurucz93} or LINFOR, and those profiles which did fit most closely, as evaluated by rms minimization, implied $(\Teff, \logg)$ solutions that disagreed significantly from those derived from photometry and metal line fitting. We suspect that because the Balmer lines span most of an echelle order, the continuum level was incorrectly estimated during normalization, so that the Balmer profiles are distorted from their true shape, and thus return an incorrect solution.

%%%%%%%%%%%%%%%%%%%%%%%%%%%%%%%%%%%%%%%%

\section{Photospheric parameter results}

\begin{figure}
\epsscale{0.70}
\plotone{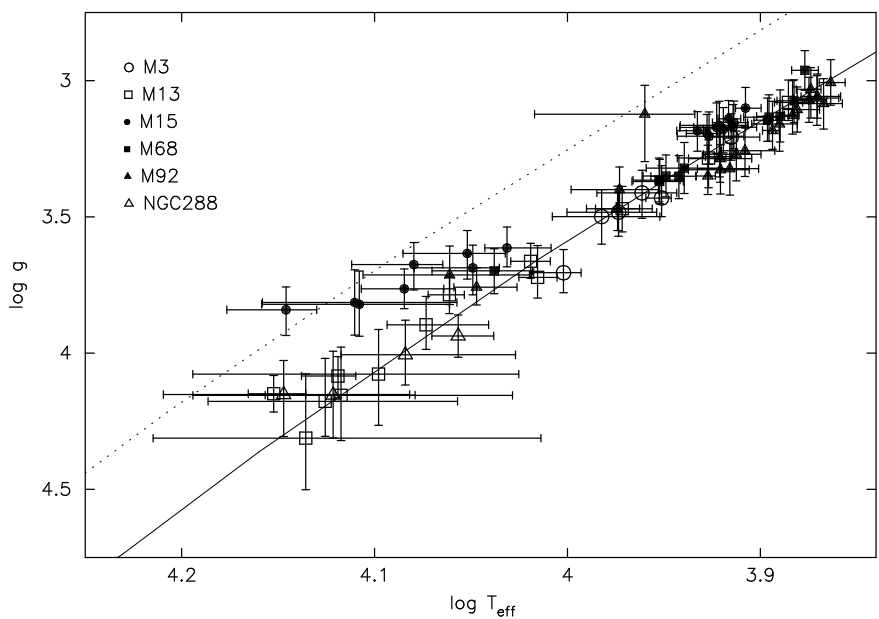}
\caption{The derived $(\Teff, \logg)$ for each of our target stars, compared to the model tracks of \cite{dorman93} for ${\rm [Fe/H]} = -1.48$. True HB stars should lie between the ZAHB (zero-age HB, solid line) and TAHB (terminal-age HB, dotted line) loci. \label{hr-bhb}}
\end{figure}

The final adopted photospheric parameters $(\Teff, \logg, \xi)$ for the program stars are listed in Table~\ref{phot-solutions}, along with an indication of which photometric data and metal species were used to constrain each parameter. To test whether these derived parameters are plausible, we plot them on a theoretical HR diagram, and compare the points to each other and to the HB locus predicted by stellar models. Figure~\ref{hr-bhb} shows points for our 74 stars, which generally lie between the zero-age HB and terminal-age HB curves from the models of \cite{dorman93}, so our analysis procedure yields parameter values which are compatible with our theoretical understanding of HB stars. 

\begin{figure}
\epsscale{0.70}
\plotone{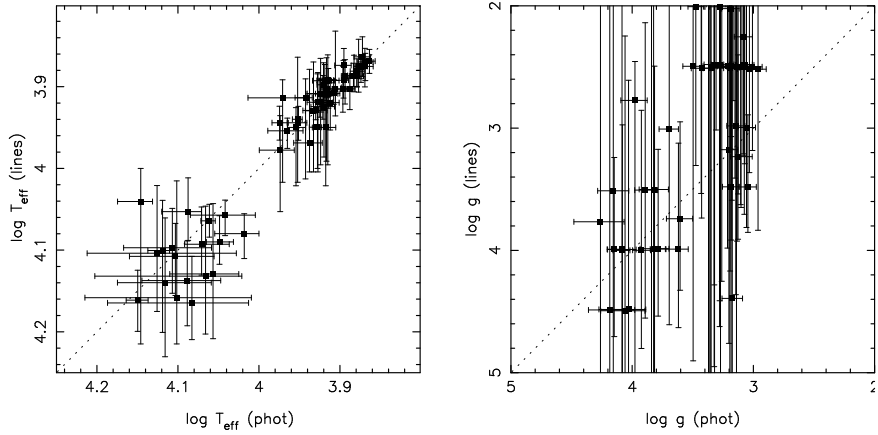}
\caption{Comparison of $\Teff$ and $\logg$ values derived from photometry alone and spectroscopy alone. The dashed lines indicate perfect agreement between the two independent techniques. The derived temperature values agree reasonably well with this line, while the error bars on the spectroscopically-derived gravities are too large to assess the quality of agreement. \label{phot-spec-compare}}
\end{figure}

Some of the stars in our sample have well-defined $\Teff$ and $\logg$ from photometry {\it and} from spectroscopic analysis of metal lines, as described in the previous section. Comparison of the results from these two separate and independent techniques gives us an opportunity to look for systematic errors in each technique. Figure~\ref{phot-spec-compare} plots the temperatures and gravities from photometry alone versus those from line analysis alone.  The temperature values adhere well to a unity relationship, and no systematic trend is evident. The error bars on the spectroscopically-determined surface gravities, unfortunately, are much too large to yield any useful information.

\begin{figure}
\epsscale{0.70}
\plotone{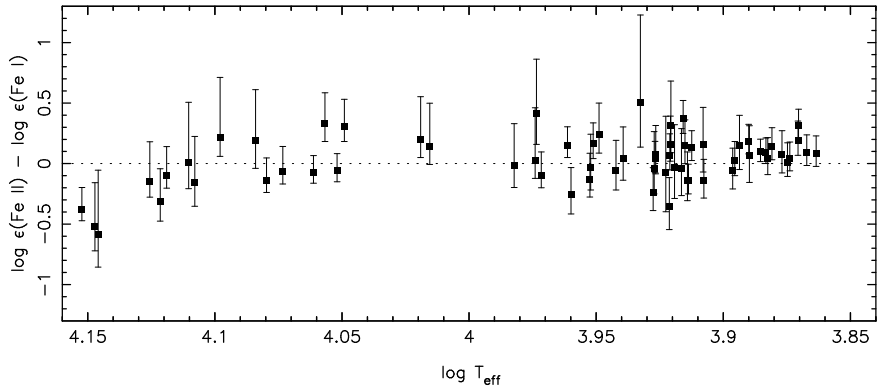}
\caption{The ionization offset, $\logeps($Fe\II$) - \logeps($Fe\I$)$, as a function of $\Teff$. Aside from a possible slight downturn at the highest temperatures, there appears to be no systematic trend in the ionization offset, and the mean value of the offset lies a few tenths of a dex above zero, as predicted by theoretical models of non-LTE corrections to LTE abundance analysis results. \label{ionz-offsets}}
\end{figure}

As a further test of the reliability of these derived values, we can look at the resulting $\logeps$ values for two different ionization stages of the same element. If we have selected the proper temperature and gravity for a given star, then we should derive the same abundance from both neutral and singly-ionized lines, assuming that our LTE models accurately reproduce the actual conditions in the photosphere of the star. For hotter, metal-poor stars, this assumption starts to break down, as various non-LTE effects, such as ``overionization'' by UV photons, come into play. Several authors have theoretically and empirically assessed the non-LTE offsets for Fe~I/II \citep{gigas86, thevenin99, allendeprieto99} and Mg~I/II \citep{przybilla01}, but most of these studies focus on stellar types rather different from our BHB stars, so we have chosen not to rely on offset-corrected ionization equilibria to constrain temperatures and gravities for this project. It is still useful, however, to see how the ionization offset varies across our sample, since a large $\logeps$ discrepancy might indicate erroneous photometric parameters for a particular star. Figure~\ref{ionz-offsets} plots $\logeps($Fe\II$) - \logeps($Fe\I$)$ as a function of derived $\Teff$. All of the stars lie within $2\sigma$ of zero offset, and most are also compatible with $\logeps($Fe\II$) - \logeps($Fe\I$) \simeq +0.2$~dex, the offset predicted by many of the studies cited above. There does appear to be a slight downturn in $\Delta\logeps$ among the hottest stars in our sample, perhaps indicating that $\Teff$ for these stars is consistently overestimated; more comprehensive photometric or spectrophotometric analysis of these stars, using bluer wavelength bands, will probably be necessary to resolve this issue.

%%%%%%%%%%%%%%%%%%%%%%%%%%%%%%%%%%%%%%%%

\section{Abundance results}

Table~\ref{eps-results} shows the derived chemical abundances $\logeps$ for each star, along with several different error quantities. The $\sigma_{\rm fit}$ columns list the internal errors (plus and minus $1\sigma$) as determined by the rms fitting, which includes contributions from the noise in the observed spectrum and discrepancies among different lines of the given species. We then calculated the change in $\logeps$ resulting from $\pm 1\sigma$ changes in $\Teff$, $\logg$, $\xi$, and $\vsini$. Errors in $\loggf$ in the catalog of atomic parameters can also influence the abundance determinations; one reason we chose the Hirata catalog was that it includes estimates of the error in $\loggf$ for most atomic transitions. Summing the $\sigma(\loggf)$ values for the specific lines used in the $\logeps$ calculation for each star, and dividing by $\sqrt{N_{\rm lines}}$, we can estimate the potential error contribution $\sigma_{g\!f}$ in $\logeps$ due to these uncertainties. All of the positive error quantities are added together in quadrature to determine the total positive error $+\sigma_{\rm total}$, and likewise for the total negative error $-\sigma_{\rm total}$. 

\begin{figure}
\epsscale{0.70}
\plotone{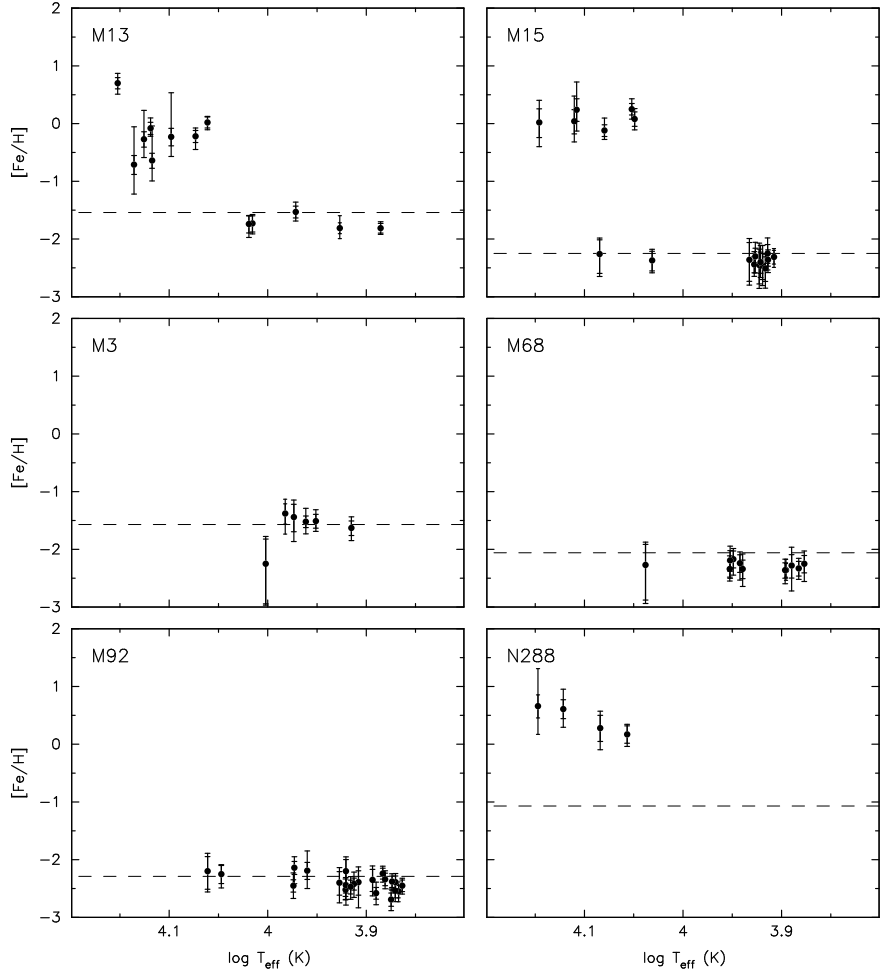}
\caption{Iron abundances [Fe/H] for program stars, plotted as a function of $\Teff$, for each cluster. Circles are abundances derived from one or more absorption lines. The inner error bars attached to each point represent internal errors, as determined by the line profile fits, while the outer error bars include the effects of variations in the adopted photospheric parameters. Horizontal dashed lines denote the canonical cluster metallicities. \label{eps-fe}}
\end{figure}

\begin{figure}
\epsscale{0.70}
\plotone{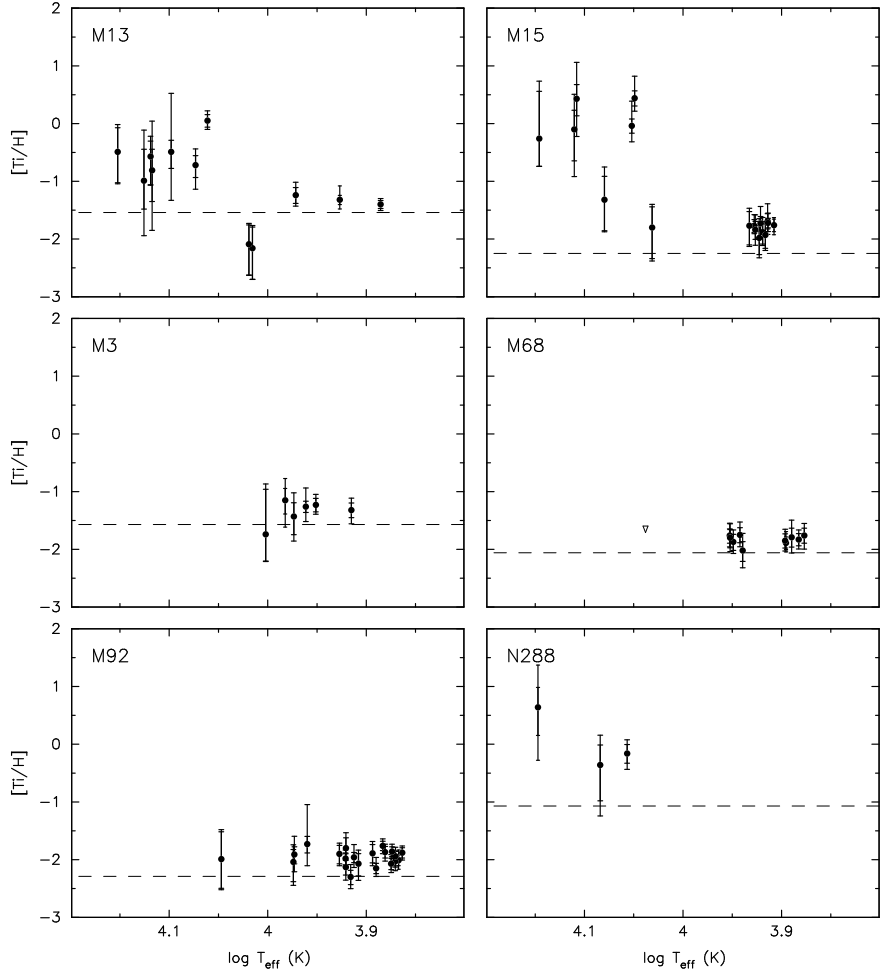}
\caption{Titanium abundances [Ti/H] for program stars. Inverted triangles indicate upper bounds on the value of $\logeps$. \label{eps-ti}}
\end{figure}

\begin{figure}
\epsscale{0.70}
\plotone{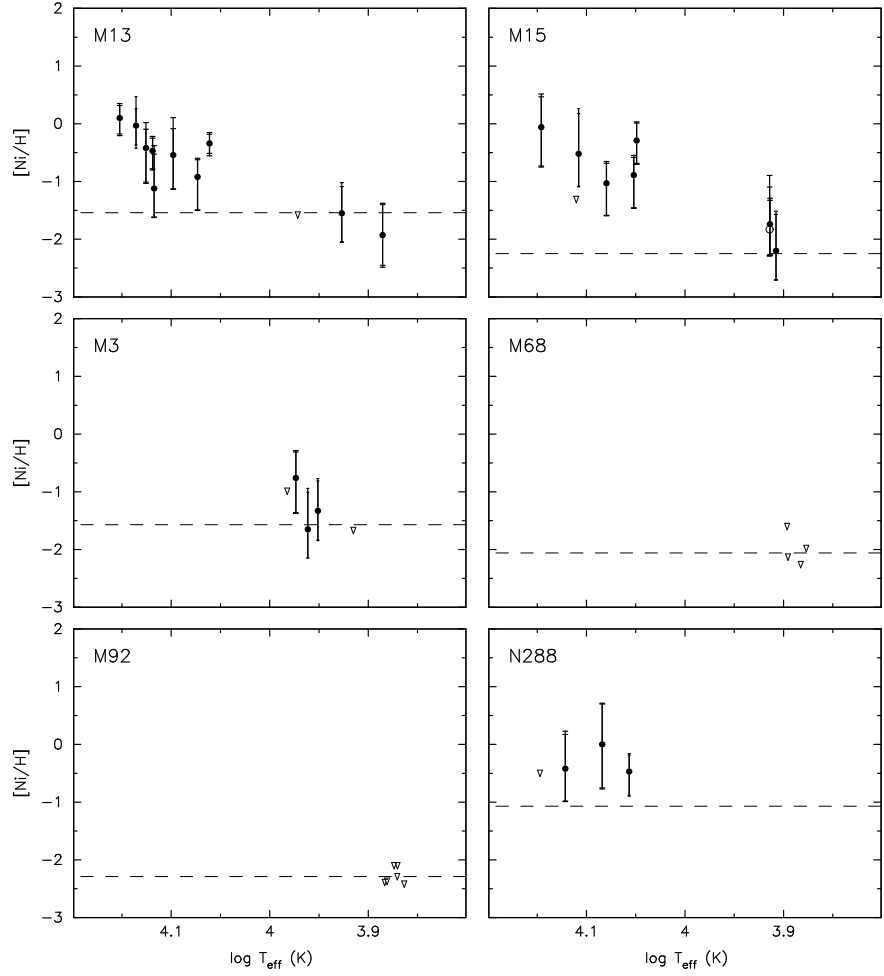}
\caption{Nickel abundances [Ni/H] for program stars. Filled circles indicate abundances derived from the dominant singly-ionized species, while open circles represent
neutral species. \label{eps-ni}}
\end{figure}

\begin{figure}
\epsscale{0.70}
\plotone{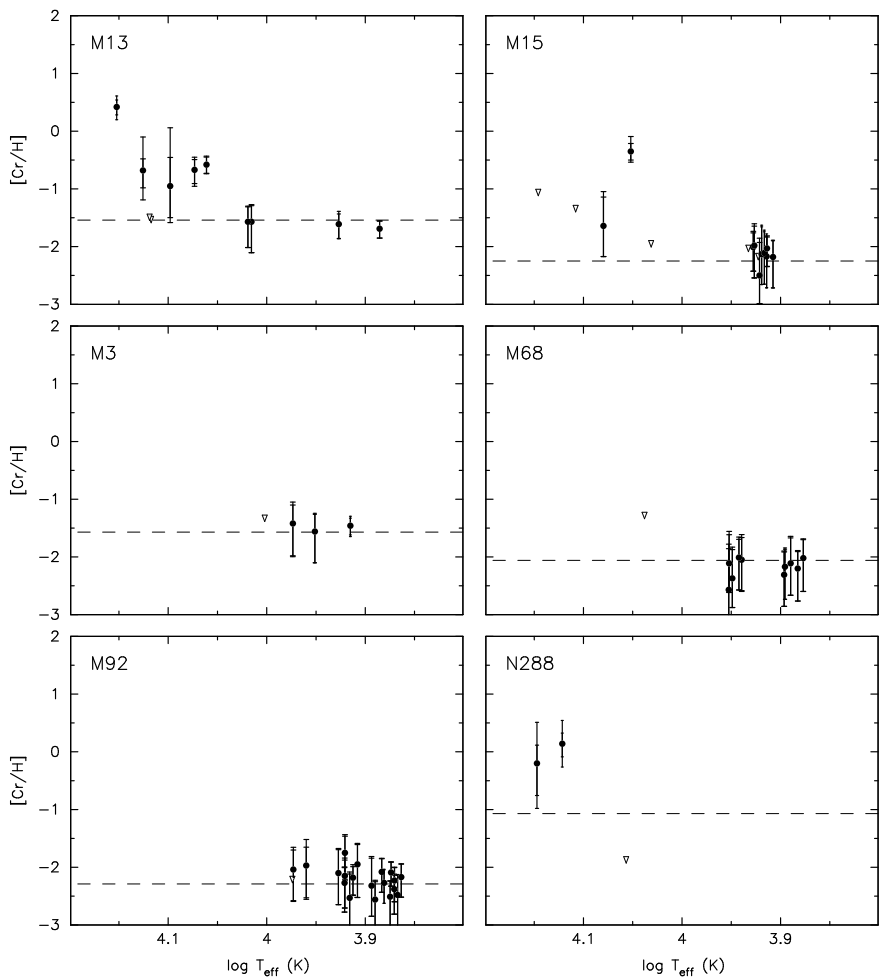}
\caption{Chromium abundances [Cr/H] for program stars. \label{eps-cr}}
\end{figure}

\begin{figure}
\epsscale{0.70}
\plotone{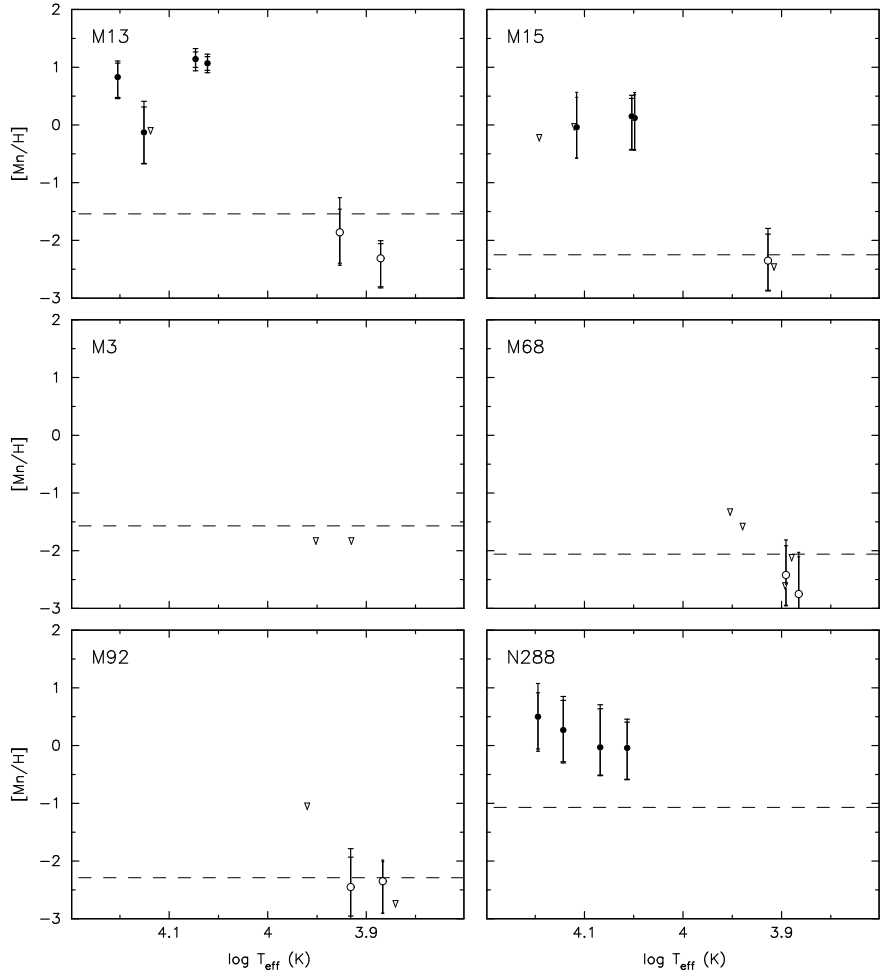}
\caption{Manganese abundances [Mn/H] for program stars. \label{eps-mn}}
\end{figure}

\begin{figure}
\epsscale{0.70}
\plotone{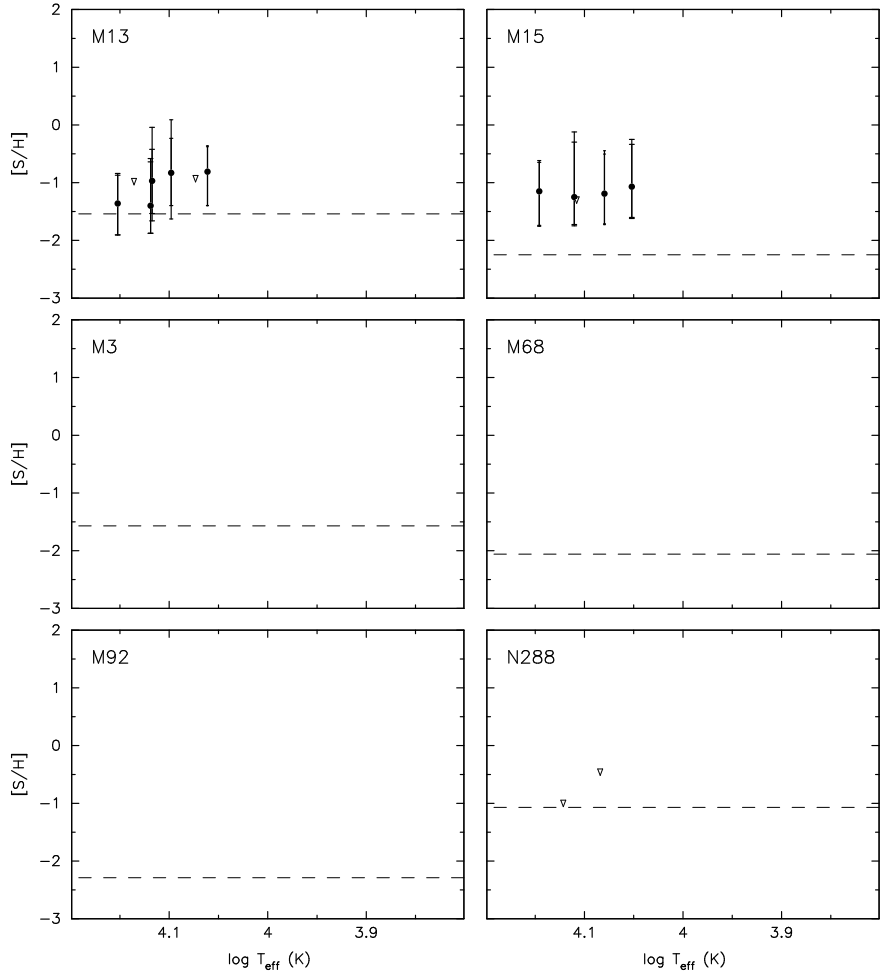}
\caption{Sulphur abundances [S/H] for program stars. \label{eps-s}}
\end{figure}

\begin{figure}
\epsscale{0.70}
\plotone{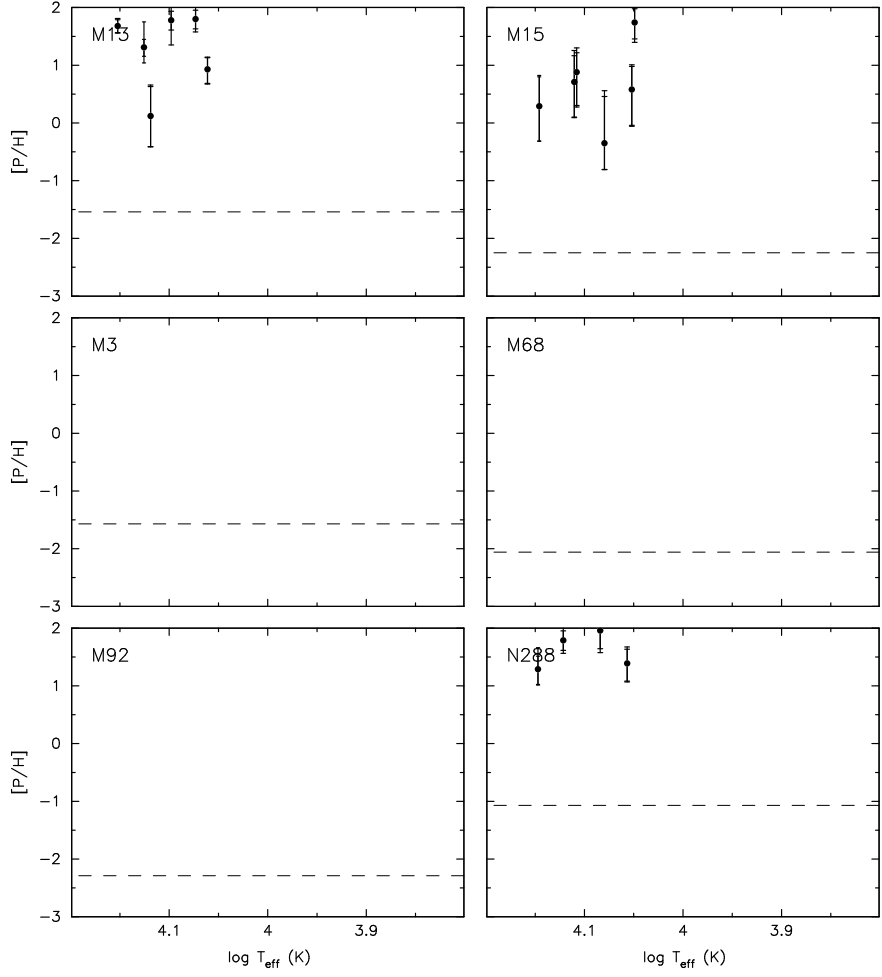}
\caption{Phosphorus abundances [P/H] for program stars. \label{eps-p}}
\end{figure}

\begin{figure}
\epsscale{0.70}
\plotone{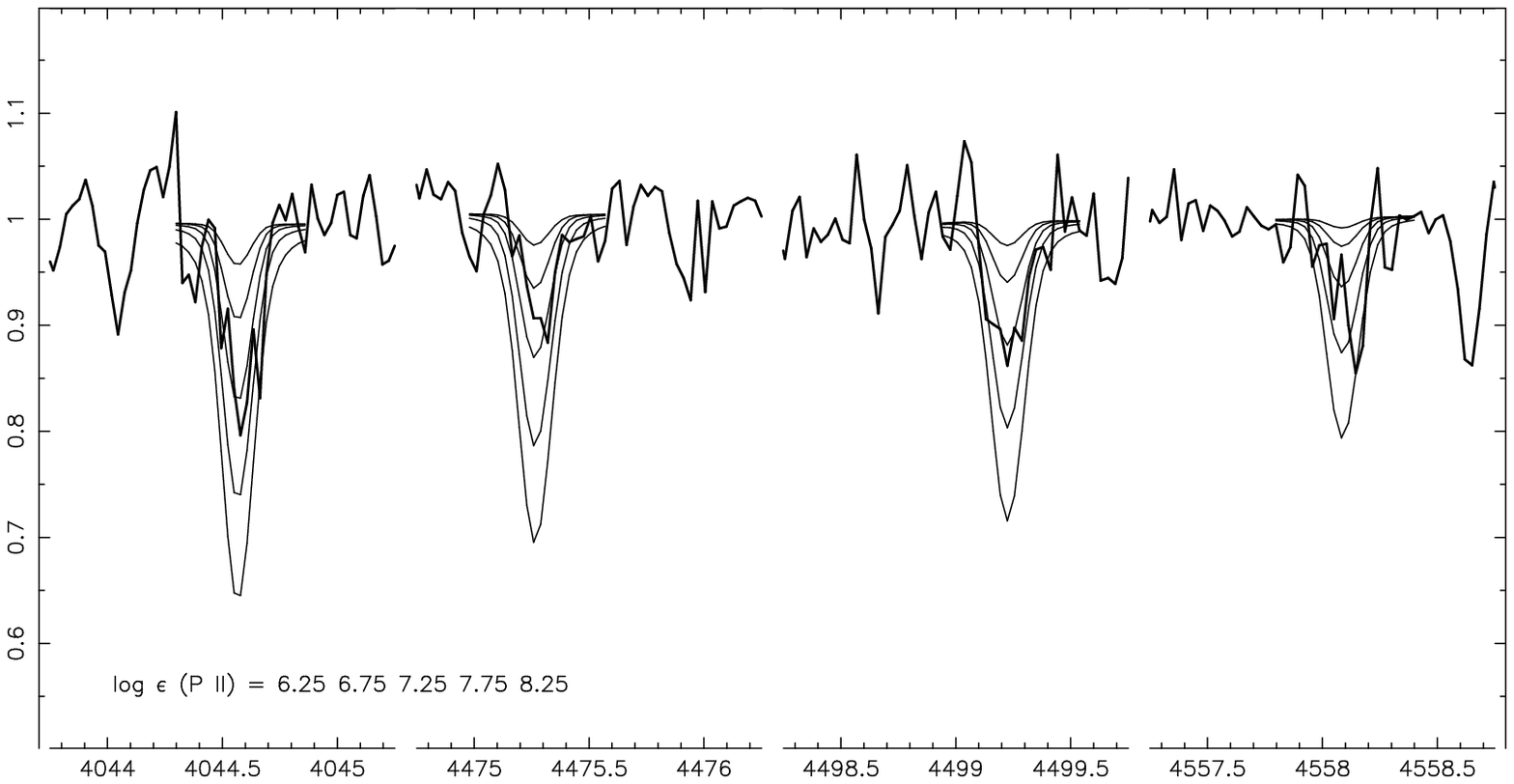}
\plotone{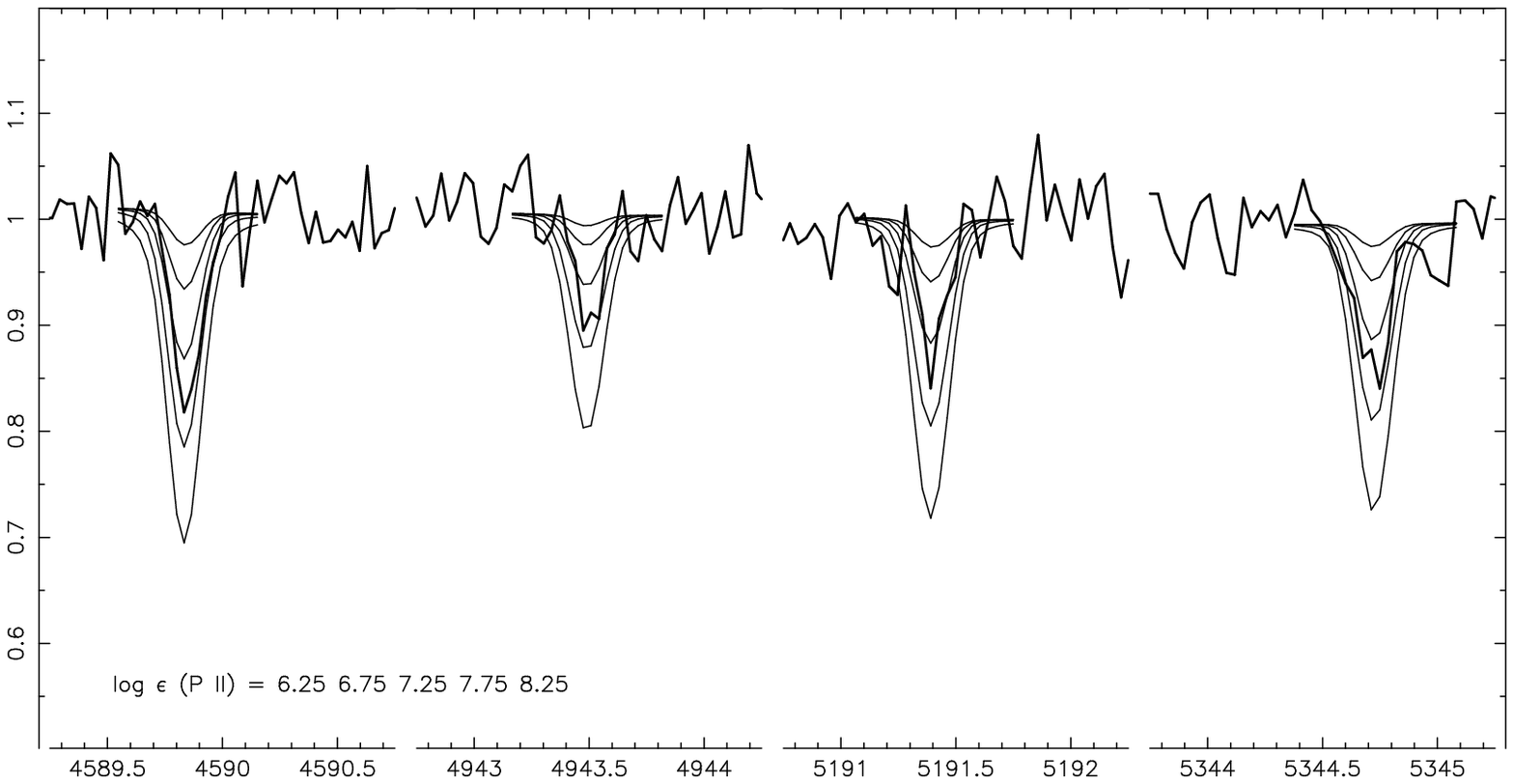}
\caption{Comparison of observed (heavy line) and synthetic (light lines) profiles for eight selected P\II\ lines of star M13/WF2-820. The adopted value of $\logeps$(P\II) runs from 6.25 ([P/H]$=+0.80$) to 8.25 ([P/H]$=+2.80$), by 0.50 dex steps. The best global fit across all lines of this species is found for $\logeps$(P\II)$=7.25$, a 1.80~dex enhancement above the solar phosphorus abundance. \label{p-synth}}
\end{figure}

\begin{figure}
\epsscale{0.70}
\plotone{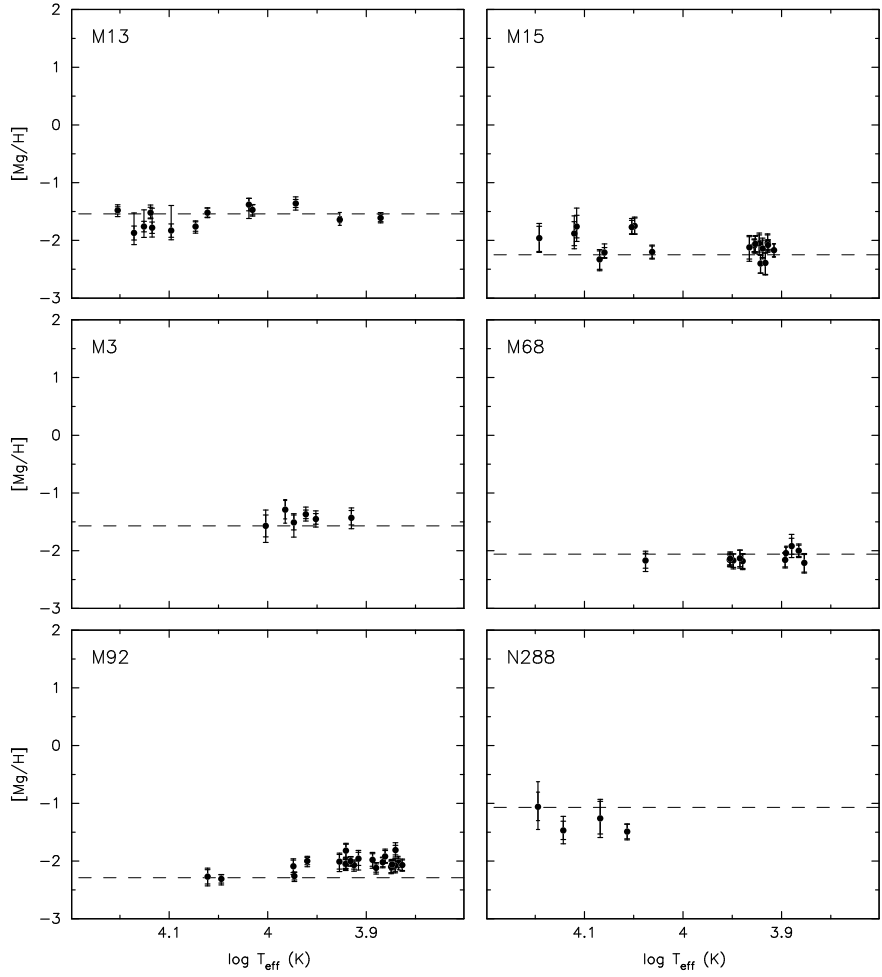}
\caption{Magnesium abundances [Mg/H] for program stars. \label{eps-mg}}
\end{figure}

\begin{figure}
\epsscale{0.70}
\plotone{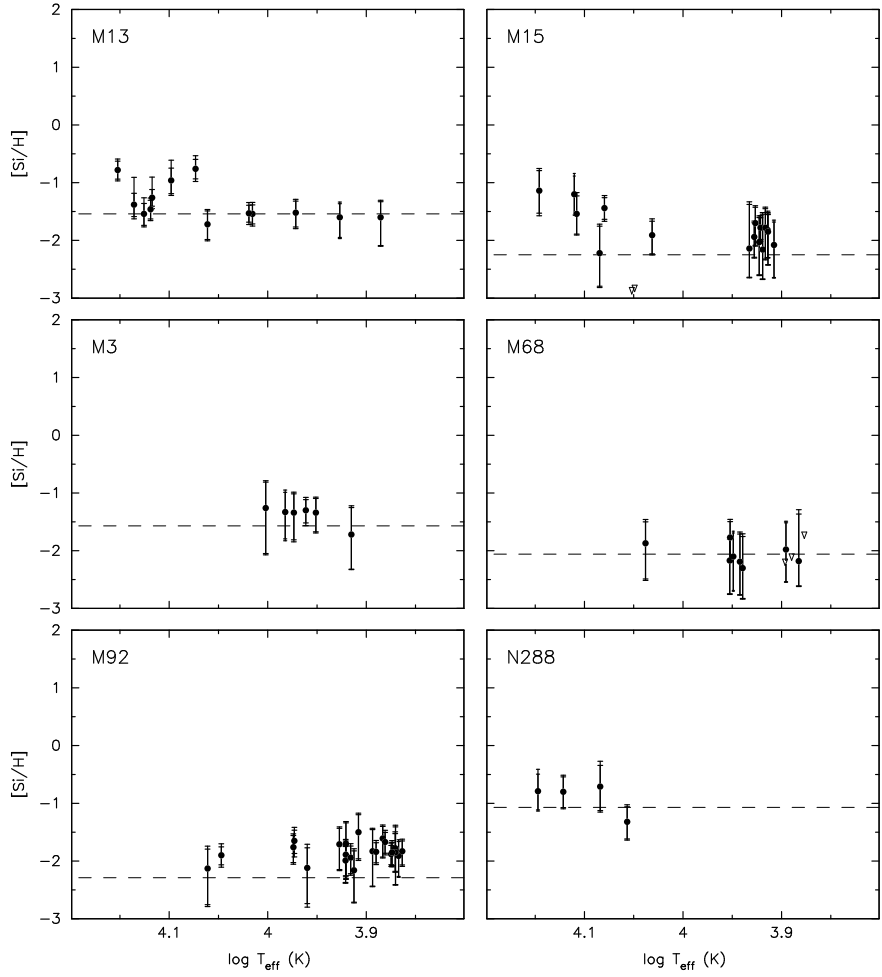}
\caption{Silicon abundances [Si/H] for program stars. \label{eps-si}}
\end{figure}

\begin{figure}
\epsscale{0.70}
\plotone{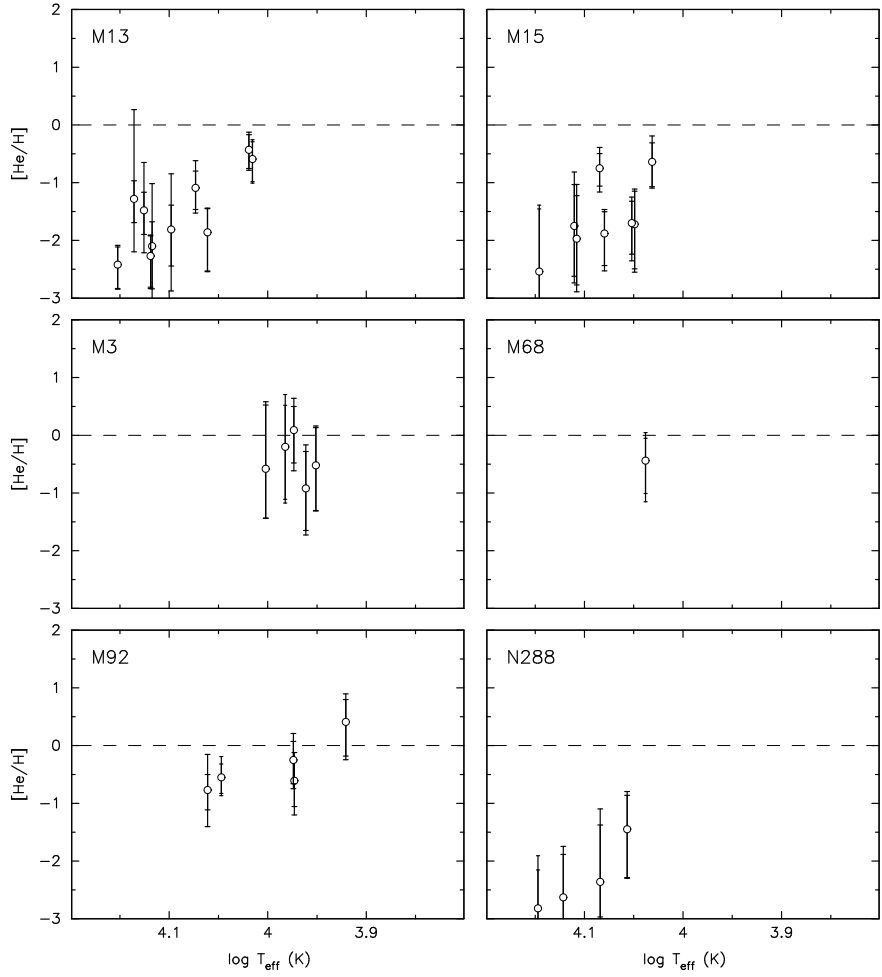}
\caption{Helium abundances [He/H] for program stars. \label{eps-he}} 
\end{figure}

The abundance trends for ten key chemical species appear Figures~\ref{eps-fe}--\ref{eps-he}, where we plot [X/H] as a function of $\Teff$ for stars in each globular cluster. As far as possible, we use the values for the dominant ionization stage (usually singly-ionized metals) as the best indicator of the actual chemical abundance, since these stages are least susceptible to non-LTE effects. The effective temperature serves to parameterize each star's position along the HB locus, without the bolometric correction issues that would be inherent in a color diagnostic like $B-V$. Circles indicate measured abundances as logarithmic offsets from the solar proportion of each element (open symbols for neutral species, solid symbols for singly-ionized species), while inverted triangles indicate upper limits, and the horizontal dashed lines mark the canonical [metal/H] of each cluster (not incorporating element-to-element variations expected from $\alpha$-enhancement or CNO dredge-up). Each abundance point has two error bars: the smaller ones indicate $\pm 1\sigma_{\rm fit}$ (internal error only), while the larger error bars are $\pm 1\sigma_{\rm total}$ (including the effects of errors in the photospheric parameters).

Table~\ref{eps-summary} summarizes the abundance results for helium, magnesium, phosphorus, titanium, and iron. Along with each best-fit [X/H] abundance value, we quote the $\pm 1\sigma_{\rm total}$ values,  and in parentheses, the number of absorption lines used for the abundance determination.

We find remarkable enhancements of iron and other metal species among the BHB stars hotter than $\sim 11200 \K$. The iron abundances among the cooler stars are close to their respective cluster metallicities, in some cases a few tenths of a dex below the [Fe/H] derived from analysis of red giant stars in each of these clusters, as seen in Figure~\ref{eps-fe}. Most of the hot stars, however, show iron content similar to that of the sun, ${\rm [Fe/H]} \simeq 0.0$. Depending on the intrinsic metallicity of the cluster, these values represent enhancements of factors of 30 to 300. Titanium, similarly, is found a few tenths of a dex above the cluster baseline in the cooler stars (such $\alpha$-enhancement is common in metal-poor stars), but then rises by factors of 10 to 100 in the hotter population (Figure~\ref{eps-ti}), although the ``step-function'' in $\logeps$ is not as pronounced as with iron. Nickel is enhanced to just below solar levels (Figure~\ref{eps-ni}). Chromium (Figure~\ref{eps-cr}) and manganese (Figure~\ref{eps-mn}) lines appear in only a few of the hot stars, but there is a clear overabundance of these elements, of similar magnitude to iron, as compared to the cool stars. Sulphur appears to be marginally enhanced among the hot stars (Figure~\ref{eps-s}).

Phosphorus exhibits significantly larger enhancements than iron (Figure~\ref{eps-p}). We do not observe any P\II\ lines among the cooler stars, but if we assume an appropriately-scaled solar composition for these stars, then the ${\rm [P/H]} \simeq +1.5$ that we find for the hot stars implies, in some cases, an enhancement of 3.5 orders of magnitude. These values are each based on several separate spectral lines (Figure~\ref{p-synth}), in close agreement with each other, so we are confident that they are not due to random errors or line misidentification. We note that a plethora of P\II\ lines was found in the field HB stars Feige~86 and 3~Cen~A by \cite{sargent67} and \cite{bidelman60}, respectively, suggesting that the same mechanism may be at work in both cluster stars and field stars.

Magnesium, on the other hand, shows little (if any) enhancement (Figure~\ref{eps-mg}). The [Mg/H] abundances that we derive are in close agreement with the canonical metallicities of each cluster, even bluewards of $11200 \K$. There are hints of a slight rise in the magnesium abundance at the highest $\Teff$ in M15, but any such enhancement is small compared to the other metal species, and may be attributable to incorrect $\Teff$. Silicon behaves in almost exactly the same fashion (Figure~\ref{eps-si}), with hints of a slight rise, no more than 0.5~dex on average, among the hottest stars of M13 and M15.

This pattern of overabundant iron and normal magnesium that we observe in our stars matches the prior results of \cite{glaspey89}, and the recent work by \cite{moehler99} on the southern metal-poor globular NGC~6752. Moehler {\it et al.} observe 42 BHB stars at medium spectral resolution, and derive [Fe/H] and [Mg/H] from the strongest lines of each species. They find a very similar jump in the iron composition for 19 stars above $\Teff \simeq 11500 \K$, but no appreciable change in the magnesium abundance. Their spectra are not of sufficient resolution or $S/N$ to assess the behavior of other metal species, but the iron and magnesium results provide an exciting parallel to our findings.

The helium composition of our stars also varies as a function of position along the horizontal branch, as seen in Figure~\ref{eps-he}, although the true size and nature of this variation is difficult to determine given the large uncertainties in most of the abundance values. Helium lines are not visible in the coolest of our stars, but at $\Teff \simeq 9000 \K$, the photospheres become hot enough to excite the He\I\ transitions at 4471, 5016, and 5876 \AA, and we find $\logeps({\rm He}) \simeq 11$, as expected for the primordial helium fraction. Towards hotter $\Teff$, however, the mean abundance of helium is lower, reaching a depletion of 2.5 dex, or a factor of 300, in some cases. Earlier analyses of stars in M13 and M15 \citep{behr00a, behr00b} claimed to see a monotonic decrease in [He/H] with increasing $\Teff$, but this trend is less evident in the updated analysis, and may not actually be present, particularly in light of the helium abundances measured for hot BHB stars in NGC~6752 by \cite{moehler00}, which show a roughly constant $\log({\rm He/H})$ over a wide span of $\Teff$.

%%%%%%%%%%%%%%%%%%%%%%%%%%%%%%%%%%%%%%%%

\section{Rotation results}

Table~\ref{vsini-results} lists the projected rotation velocities that we measure for our target stars. The error quantities $\pm\sigma_{\rm fit}$ indicates the confidence interval returned by the rms fitting routine. In addition to these internal errors, there are several potential sources of external or systematic error. Other stellar velocity fields, as due to microturbulence, macroturbulence, and thermal motions, can also contribute to line broadening. The error quantities $\pm\sigma_{\xi}$ shows how much the best-fit value of $\vsini$ varies when microturbulence $\xi$ is increased and decreased by $1 \kms$. A similar test, adding and subtracting an additional $1 \kms$ of macroturbulence to the synthetic spectra, yields $\pm\sigma_{\rm macro}$. This value is characteristic of convective motions in the Sun, and similar velocities might be expected in the cooler stars, although the hotter stars in our sample are expected to have fully radiative envelopes, with no convective or macroturbulent motions at all. Changes in thermal broadening as a result of errors in adopted $\Teff$ will be negligible ({\it e.g.} $\pm 0.1 \kms$ for $\Teff = 12000 \pm 1000 \K$ for iron atoms). If our choice of the limb darkening parameter is in error by 0.1, this will result in a slight change $\sigma_{\rm limb}$ in the best-fit $\vsini$, as the wings of the rotational broadening profile are given more or less weight. Instrumental and observational effects might also broaden the lines. HIRES, like any spectrograph, will show slight differences in instrumental profile across the detector, but by comparing median-filtered line spread functions from the arc emission lines in the four separate quadrants of the detector, we estimate such variation in the FWHM of our instrumental broadening function to be less than $0.2 \kms$. Because HIRES is located at Keck's Nasmyth focus, and does not tilt in altitude, there are no concerns about gravitational flexure, and the maximum change in observed radial velocity resulting from Earth's rotation will be $0.13 \kms$ over 1 hour, the typical length of observation for one star. The dominant instrumental effect is likely to be the ``underfilling'' of the slit by the stellar seeing profile, as compared to the thorium-argon arc lamp, which was discussed in Section~2. To evaluate the potential systematic influence of underestimated or overestimated instrumental profile widths, we repeat the $\vsini$ analysis using instrumental profiles that are 10\% wider and 10\% narrower than those initially derived, and show the resulting change in best-fit $\vsini$ as $\pm\sigma_{\rm ip}$. Finally, measurement of the width of a line will depend upon the adopted continuum level: if the continuum is set higher, then the line will appear broader to the fitting algorithm, while a lower continuum will make a line appear narrower. We force the continuum level of each spectrum up and down by 1\%, and tabulate the change in $\vsini$ as $\pm\sigma_{\rm cont}$. Adding all these errors together linearly, we get an estimate of the total systematic error, $\sigma_{\Sigma{\rm sys}}$.

\begin{figure}
\epsscale{0.90}
\plotone{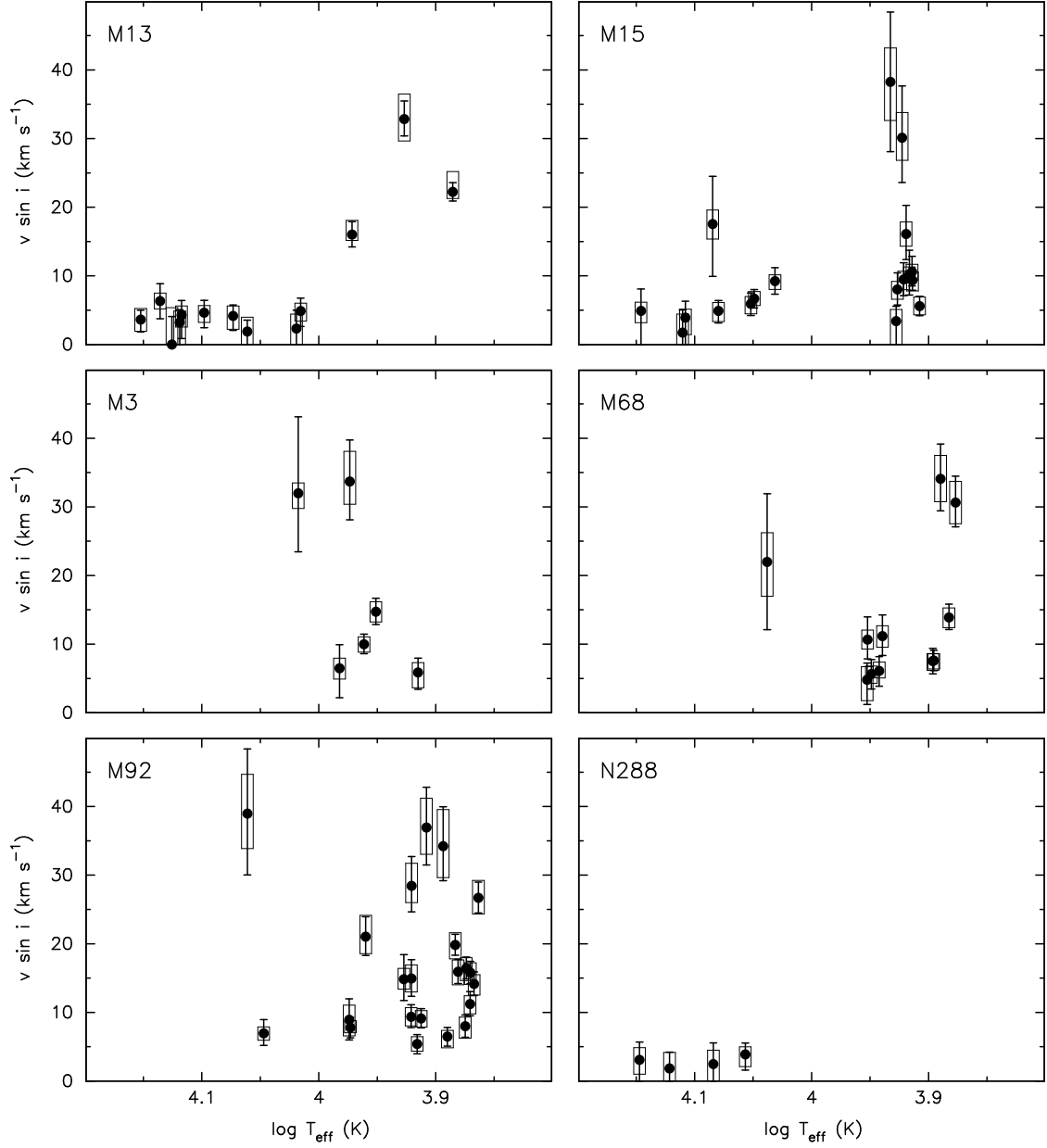}
\caption{Rotation velocities $\vsini$ as a function of $\Teff$ for stars in each of the six clusters. The traditional error bars show the random error evaluated from the quality-of-fit curve, while the narrow rectangles show the sum of various systematic errors, as described in the text. \label{vsini-results-fig}}
\end{figure}

\begin{figure}
\epsscale{0.70}
\plotone{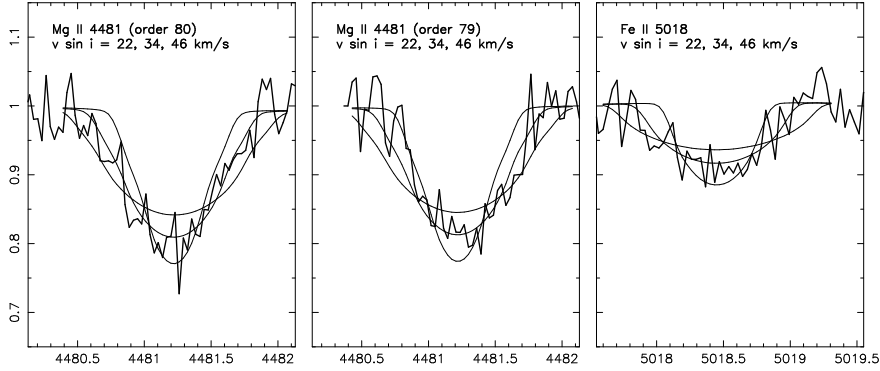}
\caption{Spectral synthesis fits for strong metal lines of the star M3/B244. For each adopted $\vsini$ value, $\logeps$ for Mg\II\ and Fe\II\ are adjusted to achieve the best fit. The optimal global fit is found for $\vsini \simeq 34 \kms$, while values of $22 \kms$ and $46 \kms$ (roughly $\pm 2\sigma$ from the best-fit value) yield less satisfactory agreement between the observed and synthetic line profiles. \label{m3-rot-1}}
\end{figure}

\begin{figure}
\epsscale{0.70}
\plotone{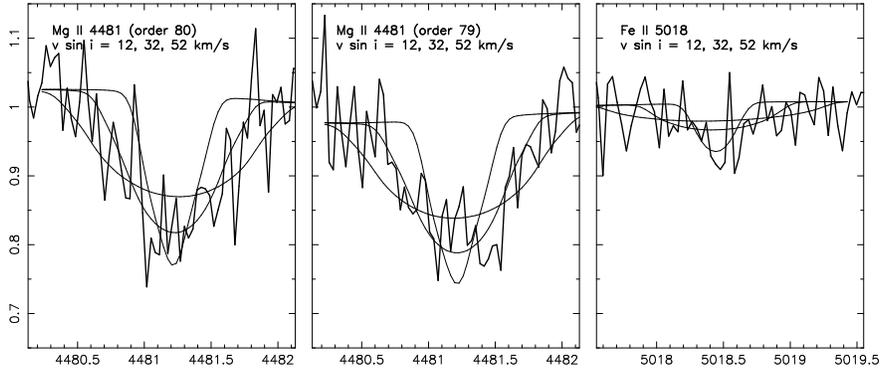}
\caption{Spectral synthesis fits for strong metal lines of the star M3/B446, similar to those shown in Figure~\ref{m3-rot-1}. A value of $\vsini = 32 \kms$ gives the best global fit across all lines, while $\pm 2\sigma$ values of $12 \kms$ and $52 \kms$ do not fit as well. \label{m3-rot-2}}
\end{figure}

Figure~\ref{vsini-results-fig} displays these rotation results in a fashion similar to the abundance plots. The vertical error bars show only the statistical error $\sigma_{\rm fit}$, as measured from the $z$-curves, while the rectangles indicate the sum of all systematic errors, $\sigma_{\Sigma{\rm sys}}$, as described above. Of particular note are two stars in M3 with $\vsini > 30 \kms$, M3/B244 and M3/B466. Previous measurements of this cluster by \cite{peterson95} found a maximum $\vsini$ of $20 \kms$, suggesting that no ``fast-rotating'' BHB stars existed in M3, but it now appears that there may exist a small population of fast rotators. It is odd that we would find this subpopulation in our much smaller sample, and the rotation measurements for both of these stars are based upon only a few weak absorption lines, but as shown in Figures~\ref{m3-rot-1} and \ref{m3-rot-2}, the line profiles are clearly broadened beyond $20 \kms$, so these $\vsini$ values are probably secure.

\begin{figure}
\epsscale{0.70}
\plotone{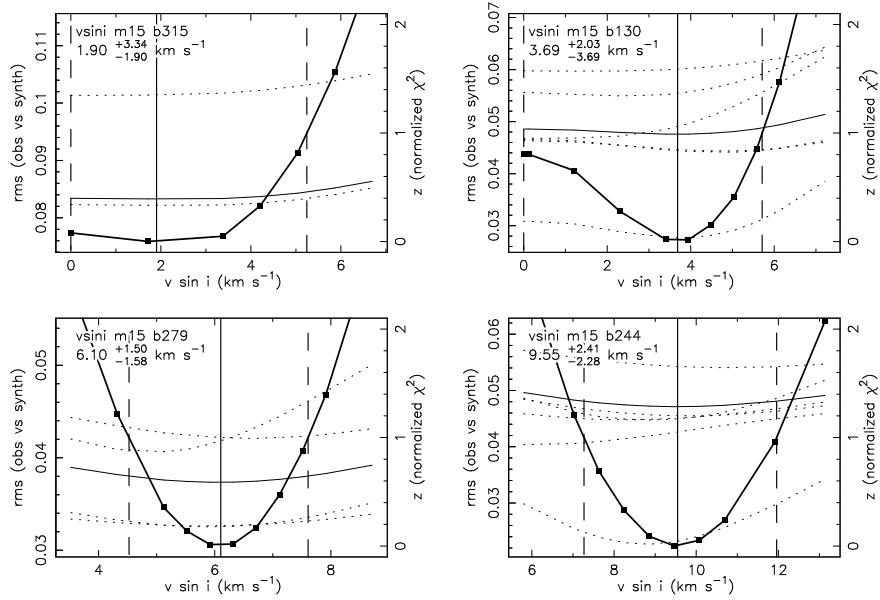}
\caption{Quality-of-fit curves for four narrow-lined stars, to illustrate detections and upper bounds on rotational broadening. The dotted curves show how the rms deviation between observed and synthetic spectra varies as a function of $\vsini$ for each separate chemical species, while the thin solid curve shows the total rms for all species combined. This composite rms is converted into a parameter $z$ (heavy solid curve), similar to a normalized $\chi^2$ measure, which is defined such that $z=0$ at the best-fit value of $\vsini$, where the composite rms reaches a minimum (vertical solid line), and $z=1$ at the $\pm 1\sigma$ error interval in $\vsini$ (vertical dashed lines). \label{rot-zcurves}}
\end{figure}

The detection of rotational broadening among the slowest-rotating stars deserves some additional discussion. Figure~\ref{rot-zcurves} shows the $z$-curves for four narrow-lined stars. For M15/B315 (upper left panel), we find a best-fit value of $\vsini = 1.90 \kms$, although $\vsini = 0 \kms$ yields an (almost) equally good fit, as the rms and $z$ values are only marginally higher at $0 \kms$ than at $1.90 \kms$, so the quoted solution of $\vsini = 1.90^{+3.34}_{-1.90} \kms$ is best considered as a $1\sigma$ upper bound of $\vsini < 5.24 \kms$. Star M15/B130 (upper right panel) has a more clearly-defined minimum in its $z$ curve, although $\vsini = 0 \kms$ cannot quite be excluded at the $1\sigma$ level. The other two stars (lower panels) have solid detections of rotational broadening, with $\vsini = 0 \kms$ firmly excluded, as $z \gg 1$ at $\vsini = 0 \kms$.

\begin{figure}
\epsscale{0.50}
\plotone{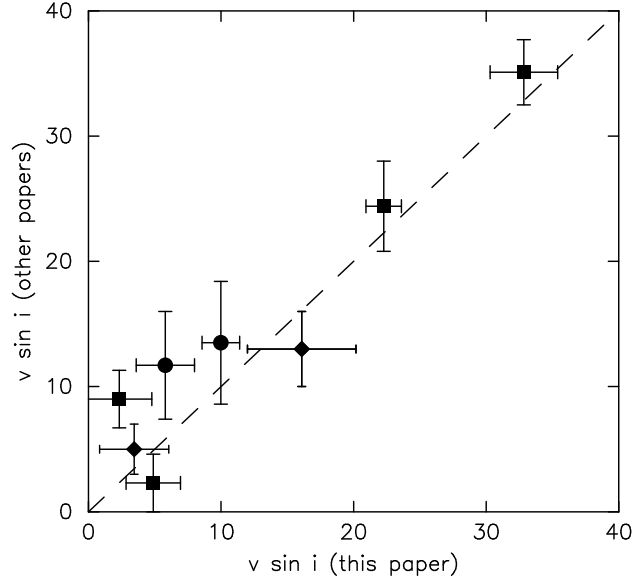}
\caption{A comparison of $\vsini$ values for stars observed by two separate observational programs. Squares are \cite{peterson95} targets in M13, circles are Peterson targets in M3, and diamonds represent \citep{recioblanco02} stars in M15. The dashed line indicates the locus of perfect agreement. \label{vsini-cmp}}
\end{figure}

A handful of our stars --- four in M13, and two in M3 --- are in common with the Peterson target list, and another two stars in M15 were also observed by \cite{recioblanco02}. The independent assessments of $\vsini$ appear to agree well within the quoted errors. In order to check for any systematic difference between our results and those of the other studies, we plot $(\vsini)_{\rm other}$ versus $(\vsini)_{\rm Behr}$ in Figure~\ref{vsini-cmp}. The points appear to agree reasonably well with a unity relation, so these three data sets thus appear to be directly comparable.

The measurement of rotation velocity via Doppler line broadening is inherently statistical, since the polar axis of each star is oriented randomly in space, and introduces an unknown $\sin i$ term to the measured quantity. Techniques do exist to deduce the inclination angle or rotation period of a star directly, using Doppler imaging or timing of the transit of surface features, but most of our stars would not be expected to show starspots or other spatial surface variability, and the combined dispersion and $S/N$ requirements would be prohibitive in any event. The only way around the unknown $\sin i$ for each star, then, is via a large statistical sample, such that we can assume an isotropic distribution of polar axis orientation, and a resulting probability function:
\begin{displaymath}
P(\sin i) \, d(\sin i) = \sin i \, \left(1-\sin^2 i \right)^{-1/2} \, d(\sin i) .
\end{displaymath}
With several stars at the same position along the HB axis, we can use this probability distribution to assess how a set of measured $\vsini$ translates into a value or range of $\vrot$.

Qualitatively, the distribution of $\sin i$ is such that large values are more likely than small values. This is convenient, as it means that a single $\vsini$ measurement is not a terribly bad estimate of the true $\vrot$ of the star. The probability that $\sin i < 0.5$, for instance, is only 0.13, and $\sin i < 0.25$ happens only 0.03 of the time. A small $\vsini$ value generally implies a small $\vrot$, therefore, and several small $\vsini$  measurements strongly limit the likelihood of a large underlying $\vrot$.

Quantitatively, we can use a Kolmogorov-Smirnov (KS) test to compare a set of observed $\vsini$ values to an assumed underlying distribution of $\vrot$. We cycle through a range of $\vrot$ values, and test the analytic cumulative probability distribution for a unimodal or multimodal rotation population against a cumulative distribution derived from the empirical measurements, using the KS parameter $P$ \citep{press92} to indicate the likelihood that the observed values were drawn from the given distribution. We treat the metal-poor and metal-enhanced populations of each cluster separately, and assess each of three cases: (1) the underlying population has a Gaussian distribution of rotation velocities with mean $v_1$ and spread $\sigma_1$ ({\it i.e.} $v_1 \pm \sigma_1$), (2) a fraction $f_1$ of the underlying population rotates at $v_1 \pm \sigma_1$, and the remainder $f_2 = 1-f_1$ rotates at $v_2 \pm \sigma_2$, or (3) a fraction $f_1$ rotates at $v_1 \pm \sigma_1$, a fraction $f_2$ rotates at $v_2 \pm \sigma_2$, and the remainder $f_3 = 1-f_1-f_2$ is at $v_3 \pm \sigma_3$. To keep the number of model parameters tractable, we assume that the Gaussian spread of each subpopulation is proportional to the mean velocity, and test five different values of $\sigma/v$: 0.01, 0.03, 0.05, 0.10, and 0.20. We step through the parameter space $(v/\sigma, v_1, f_1, v_2, f_2, v_3, f_3)$ to find the maximum values of $P$ for each of the unimodal, bimodal, and trimodal hypotheses above. For M3, M13, and NGC~288, we include the $\vsini$ values reported by \cite{peterson95} in order to create larger statistical samples. 

The results of the KS tests are compiled in Table~\ref{ks-distributions}. Values of $P$ greater than $\sim 0.1$ suggest reasonable agreement between the modal model and the observed distribution of $\vsini$. For the cooler populations, bimodal distributions are required to fit the data, as previously surmised from qualitative assessments. This preference for bimodality persists even when the Gaussian spread $\sigma/v$ is large. Trimodal models sometimes provide slightly better fits than the bimodal models, but the improvement in $P$ is not statistically significant. (No trimodal hypothesis was applied to the hot metal-enhanced population of NGC~288, because the sample size was too small to return meaningful results.) For the hot metal-enhanced populations, a single $\vrot$ is all that is required to reproduce the observed range of $\vsini$, as long as $\sigma/v$ is sufficiently large. These statistical tests provide a useful means for confirming $\vrot$ bimodality among many cool BHB populations, of estimating the proportion of fast vs. slow rotators in a cluster, but the sample sizes are still too small to distinguish the details of the underlying distributions of $\vrot$.

%%%%%%%%%%%%%%%%%%%%%%%%%%%%%%%%%%%%%%%%

\section{Explaining the abundance variations}

The large metal enhancements and helium depletion that we observe in hot BHB stars may seem startling at first, given that globular clusters are supposed to be chemically homogenous systems, but there are good observational and theoretical reasons to expect such behavior. Underabundances of helium have been observed previously in many hot, evolved stars, including subdwarf B (sdB) stars in the field and the analogous EHB stars in globular clusters \citep{baschek75, heber87, moehler00}. \cite{michaud83}, building on the original suggestion by \cite{greenstein67}, explained these underabundances as a result of gravitational settling of helium. If the outer atmosphere of the star is radiative, without convection or other large-scale flows to keep it well-mixed, then the helium atoms, with a larger mean molecular weight than the hydrogen, will diffuse downwards into the stellar interior under the influence of gravity. The actual helium content of the star remains unchanged, but the line-forming layers of the photosphere become depleted, and the He absorption lines will appear weaker than otherwise.

The Michaud \etal\ calculations indicated that helium depletion should be accompanied by photospheric enhancement of heavier elements, as the same stable atmosphere which permits gravitational settling also permits radiative levitation of metals. Elements which present sufficiently large cross-sections to the outgoing radiation field will experience radiative accelerations greater than gravity, and will diffuse upwards, enriching the photosphere at the expense of the interior. The models suggested that overabundances of factors of $10^3-10^4$ from a star's initial composition could be supported by radiation pressure, in the absence of any competing mixing mechanisms or selective mass loss. \cite{glaspey89} cite this mechanism to explain their iron-enhanced hot BHB star in NGC~6752, while \cite{grundahl99} and \cite{caloi99} both propose that such metal enhancements might be responsible for the photometric anomalies observed along the HB locus.

More recent diffusion simulations, using updated opacity tables, support this hypothesis. \cite{turcotte99} and \cite{richer00} modelled the atmospheres of main-sequence A and F types in an attempt to explain the abundance patterns observed in chemically peculiar (CP) stars. Their predictions for the relative amounts of enhancement/depletion are illustrated by Figure 11 of Richer {\it et al}. For a slowly-rotating main sequence star with temperature and gravity roughly analogous to our warmer BHB stars, their predicted abundance patterns match our empirical findings surprisingly well, albeit with much smaller magnitudes of $\Delta\log\epsilon$. Iron is enhanced by 0.7~dex in their models, titanium by 0.3~dex, and phosphorus by 0.8~dex --- a qualitatively similar pattern to the metal enhancements that we observe in our hot stars. The radiative levitation force on magnesium appears to be closely balanced with gravity, as [Mg/H] in the models is largely unchanged, which would explain the observed invariance of this element in our stars, and helium is depleted in the simulations, by 0.2 to 0.3~dex. These models will hopefully soon be applied to the specific case of metal-poor horizontal branch stars, to verify that the abundance variations that we observe can be quantitatively explained by the diffusion scenario.

A complete model of BHB star abundance variations must also explain why the metal enhancements appear so abruptly, such that the transition from metal-poor to metal-rich photospheres takes place over a span of only a few hundred degrees in $\Teff$. \cite{caloi99} suggests that this bifurcation is due to the disappearance of subsurface convection layers at a critical $\Teff$ threshold, and this qualitative assessment has received some important support from work done by \cite{sweigart02}. His BHB models show that although fully-convective envelopes disappear once the surface $\Teff$ rises above 8000~K, there still exist thin layers of convection, at or just below the surface, at isotherms corresponding to hydrogen and helium ionization. As the photospheric temperature increases, these isotherms move closer to the surface, until $\Teff \simeq 12000 \K$, when the last of these convection layers disappears, and the atmosphere is fully radiative. Since convection is highly efficient at preventing diffusion from operating, the presence or absence of even modest amounts of subsurface convection may be the ``switch'' that regulates the appearance of the metal enhancements and helium depletion. For models with $\Teff > 20000 \K$, in the EHB regime, another thin convection zone driven by helium ionization approaches the surface, possibly remixing the atmosphere and preventing diffusion from occurring. Abundance measurements of such hot EHB stars are still quite limited, so we cannot test this hypothesis directly, but the apparent disappearance of anomalously low gravities and anomalously high $u$ magnitudes (see below) suggest that this remixing may indeed be taking place.

One additional factor which might influence the diffusion mechanisms is the rotation rate of the star. This issue was approached from a theoretical standpoint by MVV. In their models, rotation velocities of a few tens of $\kms$ induced meridional circulation currents of sufficient magnitude to prevent the appearance of element diffusion. \cite{behr00b} noted that a handful of stars in M15, M68, and M92, which appear to belong to the hot BHB population, had higher rotation velocities and normal cluster abundances, and claimed that these ``anomalously non-anomalous'' stars provided evidence that faster stellar rotation can prevent diffusion from altering the photospheric abundances. However, more recent modelling of meridonal circulation \citep{michaud03} suggests that the $\vrot$ threshold is approximately $80 \kms$, considerably higher than any observed BHB rotation. Furthermore, with one exception, the derived $\Teff$ values for these stars are very near the threshold temperature of $11200 \K$, so they very well may belong to the cooler BHB category, despite lying just bluewards of apparent gaps in the BHB distribution. If this is indeed the case, then the CMD gaps in these particular clusters are not due to the onset of metal enhancements, as we discuss below.

%%%%%%%%%%%%%%%%%%%%%%%%%%%%%%%%%%%%%%%%%

\section{The influence of abundances on photometry}

Since the spectral energy distribution emerging from a stellar photosphere is strongly dependent on the opacity distribution (and hence the metallicity) of the atmosphere, the large metal enhancements that we find in the hot BHB stars might be responsible for some of the photometric peculiarities --- gap G1, $u$-band overluminosities, and the low measured gravities --- described in Section~1. According to the first parameter effect, higher metal content produces redder photometric colors, as the envelope expands and the larger surface area prompts a smaller $\Teff$. However, since the observed metal enhancements apply only to the photosphere, and not the envelope as a whole, we might expect very different photometric properties from these stars. 

Some initial quantitative work in this regard has been performed by \cite{grundahl99}, who predicted that metal-enhanced atmospheres might be common among hot BHB stars, even before the abundance results of \cite{moehler99} and \cite{behr99} were published. They proposed a mechanism whereby an abrupt increase in metal-line opacity, particularly in the ultraviolet, reduces the relative contribution of hydrogen opacity. Since the $U$ and Str\"omgren $u$ bands are sensitive to the size of the Balmer jump, where hydrogen is the dominant opacity source, the increased metal opacity will actually result in {\it higher} flux in this region of spectrum, and the near-UV magnitudes will be brighter. Using ATLAS9 flux models, they estimated that a metallicity increase from [Fe/H]$ = -1.5$ to $+0.5$ would raise the $u$ brightness by 0.3~mag, similar in size to the $u$-jumps that they observed in the CMDs of many globular clusters.

\cite{moehler99} explore the possibility that such metal-enhancements and the resulting changes in the atmospheric structure might also account for the anomalous gravities. For BHB stars in the metal-poor globular cluster NGC~6752, they derive $\Teff$ and $\logg$ from model fits to observed Balmer line profiles, adopting different metallicities and amounts of helium mixing. They find a modest improvement in the agreement between the observational HR diagram and the theoretical ZAHB locus for a solar metallicity scaled by $+0.5 \unit{dex}$ and helium at $-2.0 \unit{dex}$, but the convergence is even better when the models include internal mixing of helium into the hydrogen-burning shell. (This sort of deep mixing might be expected as a result of core rotation, which will be discussed later.) 

More detailed modelling was described by \cite{huibonhoa00}, who explicitly included diffusive mechanisms in the computation of their BHB atmosphere models. They found significant differences in the photometric locus of the ZAHB, depending on the presence of diffusion-driven metal stratification in the atmosphere. Assuming that diffusion turns on ``quickly'' around 11200~K, as our abundance results imply, then the Hui-Bon-Hoa {\it et al.} models predict jumps or gaps as well: Str\"omgren $(u-y)$ suddenly decreasing by 0.2~mag, Str\"omgren $u$ brightening by 0.4~mag, and $U-V$ decreasing by 0.1~mag. Furthermore, they find that Balmer profiles computed from their stratified models yield a decrease in $\logg$ amounting to as much as 0.5~dex, perhaps enough to explain the low observed gravities of the hot BHB stars without helium mixing.

It is not clear, however, that the transition from metal-poor to metal-enhanced photospheres can explain all the CMD gaps observed in the vicinity of $10000$--$12000 \K$. Narrow, sharply-defined gaps, as found in M13, do appear to be directly associated with the Str\"omgren $u$ jump and the photospheric metallicity jump, and can be plausibly explained by a change in the atmospheric structure, as demonstrated by Hui-Bon-Hoa {\it et al}. However, wider gaps and underpopulated regions, such as those that appear in the CMDs of M15 and M92, may not be associated with these mechanisms. These gaps are considerably larger than the photometric shift expected from the observed metal enhancements. Furthermore, they are slightly redder than the $u$ jump location, and we find a few normal-metallicity stars on the blue side of the gaps in M15 and M92, instead of exclusively metal-enhanced stars, as in M13 and NGC~288. We tentatively conclude, therefore, that the $(B-V)_0 = -0.04$ gap in M15, the $(B-V)_0 = 0.05$ gap in M92, and the small separation at $(B-V)_0 = -0.03$ in M68 must be explained by some means other than the metallicity jump.

%%%%%%%%%%%%%%%%%%%%%%%%%%%%%%%%%%%%%%%%%%%%%%

\section{Explaining the rotation characteristics}

Our line broadening fits for stars in M13 confirm the large $\vsini_{\rm max} \simeq 40 \kms$ values derived by \cite{peterson95}, and we find more such fast-rotating stars in M15, M68, and M92. \cite{recioblanco02} discover a similar distribution of rotation speeds for the cool BHB stars in M79, with a peak $\vsini$ of $\sim 30 \kms$, and find an additional fast rotator in M15. It appears that fast BHB rotation, although not present in all clusters, is a fairly common feature.

It is difficult to explain why some of these BHB stars are spinning so fast. When on the main sequence, these stars were solar type or later, and are therefore expected to have shed most of their angular momentum via magnetically-coupled winds early in their main-sequence lifetimes, reaching the $\vrot \simeq 2 \kms$ observed in the Sun and other similar Population I dwarfs. Assuming solid-body rotation and a homogeneous distribution of angular momentum per unit mass, \cite{cohen97} estimate that the BHB stars should be rotating no faster than $10 \kms$. More sophisticated modeling, such as that of \cite{sills00}, find similarly modest BHB rotation rates given $\vrot < 4 \kms$ on the main sequence and reasonable assumptions about angular momentum evolution during the RGB ascent. These predictions do not conflict seriously with the measured $\vsini$ for the majority of BHB stars, but they clearly pose a challenge for the fast-rotating cool population.

One popular explanation for the fast rotation, proposed by \cite{peterson83a} and developed further by \cite{pinsonneault91}, is that magnetic braking on the main sequence only affects a star's envelope, while the stellar core retains much of its original rotation velocity. This reservoir of angular momentum persists throughout the star's main sequence lifetime, and then spins up the envelope only after the star arrives on the HB. If a fraction of GC stars possess rapidly rotating cores, then this would not only explain the BHB rotation bimodality, but might also serve as a second-parameter candidate. If the internal rotation is fast enough to provide appreciable centrifugal support, this could reduce the pressure in the inert helium core of RGB tip stars, delaying the helium flash, such that the helium core grows larger and more envelope mass is lost, resulting in bluer HB stars \citep{mengel76, sweigart98}. Core rotation might conceivably also play a role in explaining the wide range of oxygen abundances \citep{kraft97} and aluminum-sodium correlations and aluminum-oxygen anticorrelations \citep{shetrone96} which appear in some RGB stars. The observed abundance patterns suggest that deep mixing brings fusion-processed material from the hydrogen-burning shell to the surface of some stars, while other stars show no such effect. If material is being exchanged between the core and the envelope of a star, then angular momentum may also be redistributed, yielding differences in surface rotation velocity between the stars that undergo deep mixing and those that do not. Conversely, intrinsic differences in internal rotation profiles may determine which stars are deeply mixed and which are not. Comparisons of RGB abundances and BHB rotation in M3 versus M13 suggest a connection between rotation and mixing; as more clusters are investigated, it will be  interesting to see whether this cluster-by-cluster correlation between  RGB abundance anomalies and BHB rotation persists.

The core rotation hypothesis faces some difficulties, however. Helioseismology studies suggest that the solar interior rotates nearly as a solid body down to $0.2 R_\odot$ \citep{corbard97, effdarwich02}, and \cite{charbonneau98} claim that the core inwards of $0.1 R_\odot$ cannot be rotating any faster than twice the surface rate. Slow core rotations are also inferred for young stars \citep{bouvier97, queloz98}. Admittedly, these measurements of Population~I objects cover a very different metallicity regime from the HB progenitors in metal-poor globulars, so core rotation could still be possible for the GC main sequence stars --- perhaps lower envelope metal opacity results in weaker or shallower convection, and hence less coupling between core and envelope \citep{pinsonneault91}. However, if core rotation is the immediate cause of the higher BHB rotation speeds that we observe, then we would also have to explain why only some stars in some clusters possess a rapidly rotating core. Also, a direct correlation between faster stellar rotation and bluer HB morphology no longer seems as likely as once thought. \cite{peterson95} note that NGC~288, which has a very blue HB, possesses no fast rotators, contrary to their prior suggestion \citep{peterson83b} that clusters with fast BHB rotation would tend to have bluer HB distributions. More recent observations in M68 (fast rotators, but nearly equal numbers of red and blue HB stars) and NGC~2808 (blue BHB and long blue tail, but only slow rotators) appear to rule out core rotation as the sole second-parameter agent.

An internal angular momentum reservoir might still be a plausible explanation for the observed distribution of BHB rotation velocities. \cite{sills00} explore several models for the internal angular momentum evolution of a star following its departure from the main sequence, and find that core rotation can develop during the RGB stage as a natural consequence of core contraction and envelope expansion, assuming that differential rotation in the convective envelope is permitted. They also point out that the transfer of angular momentum from a faster-rotating core to the envelope may not occur immediately upon arriving on the ZAHB. If the core-envelope coupling takes place on timescales comparable to the HB lifetime of a star, then many BHB stars will still have slowly-rotating envelopes, such that we observe small $\vsini$. Only the older stars, which have been on the HB long enough that the core angular momentum has percolated up to the envelope, show fast rotation. Furthermore, this transfer of angular momentum might be slowed, or even prevented outright, by steep gradients in mean molecular weight within the stellar interior. Sills and Pinsonneault propose that the gravitational settling of helium from the photosphere to the interior of the hotter BHB stars creates just such a gradient, which prevents their envelopes from spinning up, thereby explaining the small $\vsini$ which we observe. Not only does this model conveniently explain the rotation characteristics of both the hotter and cooler BHB stars, but it also offers a testable observational result: the faster-rotating stars, being older, will have evolved upwards in luminosity away from the ZAHB. Plots of $\vsini$ vs. $V$ magnitude for the cooler BHB stars show no trend of this sort, although the available photometric data come from several different sources, so photometric errors may mask any correlation. \cite{recioblanco02} apply a similar test to their measurements, and find no correlation either. More precise and homogenous photometry, and a larger set of $\vsini$ measurements of cool BHB stars, could conclusively address this intriguing hypothesis.

Conversely, diffusion effects may help BHB stars lose angular momentum to stellar winds. \cite{vink02} model mass loss for HB stars, and suggest that higher photospheric metallicity due to radiative levitation could greatly enhance the mass loss rate, and thus the rate at which angular momentum is carried away by the stellar wind. The hotter BHB stars might thus be spun down to the small velocities that we observe, while the cooler stars, with normal metal-poor photospheres, would have much weaker winds and thus less angular momentum loss. The authors caution that we have yet to quantitatively understand the mechanism by which angular momentum is carried away by stellar winds, but if this scenario proves feasible, then the observed distribution of $\vrot$ along the BHB could still be compatible with the hypothesis that core rotation, and perhaps deep mixing, acts as a second parameter.

Another possibility for explaining the rotation bimodality is that the slow rotators are the `normal' result of single-star evolution, and that the rapid rotators represent the results of stellar mergers, perhaps a subpopulation of blue stragglers (BS), similar to the suggestion by \cite{fusipecci92} that BS progeny populate the red HB. Such stellar merger products would likely retain excess angular momentum even through the HB stage, although the issue is complicated by other possible effects of mergers, including mass loss. It also seems unlikely that a merger product would evolve in a similar fashion as a normal single star, and appear on the same HB locus. It might still be potentially worthwhile to undertake a numerical comparison of the fast-rotator and blue straggler populations of these clusters, but such a study is beyond the scope of this paper.

In a similar vein, some recent papers \citep{soker98, siess99, soker00, soker01, livio02} explore the possibility, previously suggested by \cite{peterson83a}, that some cluster stars spin up when they swallow substellar objects --- Jovian planets or brown dwarfs --- during the red giant phase. Although a mechanism of this sort would certainly provide plenty of angular momentum, and perhaps also enhance mass loss to form bluer HB stars, it has yet to be demonstrated that planets are common in metal-poor GCs. Nearly all of the exosolar planets found to date have host stars with significantly {\it higher} metallicity than solar \citep{gonzalez00, reid02}, and the frequency of large close-orbit planets around dwarf stars in the moderately metal-poor cluster 47 Tucanae appears to be quite small \citep{gilliland00}. These observations do not rule out the possibility of orbital companions at larger distances, or mergers with free-floating low-mass objects in the cluster, but the existence of such objects should be demonstrated via other means before they are used to explain fast BHB rotation or the second-parameter effect.

Alternatively, the fast rotators may have been spun up by non-merger interactions with another star --- either tidal locking with a binary companion, or a random tidal encounter with another single star in the cluster. \cite{peterson83a} consider various possibilities for tidal locking between an evolved star and a main-sequence companion, and find a ``physically plausible'' scenario involving synchronization of a RGB star by a companion, after which the red giant evolves to the HB. None of their fast-rotating field BHB stars showed appreciable radial velocity variation, however, and subsequent measurements of velocity dispersions of GC BHB populations \citep{peterson83b} also argue against binarity. To further test this hypothesis, we compare the measured heliocentric radial velocities of four BHB stars which were observed at two different epochs. Table~\ref{vr-compare} shows that three of these stars exhibit no statistically significant reflex motions over baselines of several years, suggesting an absence of massive companions in close orbits. The fast-rotating star IV-83 in M13, however, exhibits a significant $10 \kms$ difference in $v_r$ between Peterson's measurement in 1982/1983 and our observation in 1998, and thus deserves further attention, particularly since this star is a rapid rotator, with $\vsini = 33 \kms$. More regular $v_r$ observations of a larger sample of fast-rotating BHB stars will be necessary to properly investigate the hypothesis that binary companions are related to fast BHB rotation. Close tidal encounters with other stars in the dense cluster environment have also been suggested as a potential source of the ``excess'' angular momentum, although hydrodynamical simulations indicate that only a small subset of impact parameters will produce tidal spin-up without a collision or merger. Furthermore, there appears to be no strong correlation between cluster density (which would determine the frequency of close encounters) and the presence of fast BHB rotation; at least a few fast-rotating BHB stars are found in the sparse cluster M68 and the even sparser field population, while the dense cluster NGC~2808 does not appear to have any fast rotators.

%%%%%%%%%%%%%%%%%%%%%%%%%%%%%%%%%%%%%%%%

\section{Conclusions and future directions}

We have performed detailed abundance analyses of BHB stars in six different metal-poor globular clusters. The cooler stars, with $\Teff < 11200 \K$, show photospheric abundances which agree with the expected composition for each cluster, while the hotter stars, with $\Teff = 11200$--$20000 \K$, exhibit large enhancements of most observed metal species, and depletion of helium. Iron, titanium, nickel, chromium, manganese, sulphur, and phosphorus are enhanced by factors of 10 to 3000, while magnesium and silicon abundances are unchanged. These abundance variations are similar to those predicted by models of atomic diffusion processes --- radiative levitation of the metals, gravitational settling of the helium --- which occur in the absence of atmospheric convection. The metal enhancements appear to be capable of altering the emergent flux distribution of these stars, thus contributing to photometric peculiarities, such as luminosity and $\logg$ jumps, that appear in cluster CMDs. Metal enhancements might also cause some, but not all, of the HB gaps observed in many clusters. 

We also measure the projected rotation velocities $\vsini$ of our sample set, and again find a significant difference between the cooler and hotter stars. In some clusters, approximately one-third of the cooler stars are rotating quite fast, 30--$40 \kms$, as previously reported in the literature, while the other two-thirds have more modest $\vsini$ values of 15--$20 \kms$. The hotter stars, however, all rotate slowly, with $\vsini < 8 \kms$. Several different scenarios have been proposed to explain this distribution in rotation velocities, but it is not currently clear which of them (if any) are correct.

Differences in the fast-rotating population from cluster to cluster may provide some important clues regarding the mechanism which creates the $\vsini$ bimodality. As already noted, some clusters have fast rotators, while others do not. Furthermore, not all of the fast rotating populations appear to be the same. The highest observed $\vsini$ ranges from $40 \kms$ in M13, to only $28 \kms$ in M79, with other clusters (and the metal-poor field population) lying in the 30--$35 \kms$ range. Also, the proportion of fast vs. slow cool BHB stars may vary from cluster to cluster --- in M79, 8 of 13 cool stars have $\vsini > 15 \kms$, while in M15, only 3 of 14 stars are fast rotators. These apparent differences could be due to small sample sizes, but they could instead indicate a spin-up mechanism whose effectiveness (maximum $\vrot$ and fraction of stars affected) depends on cluster parameters. Observations of BHB rotation in many additional clusters will be necessary to determine whether the presence and nature of a fast BHB population correlates with global characteristics of the parent cluster, such as density, velocity dispersion, cluster rotation, or blue straggler population. Further $\vsini$ study of metal-poor field HB stars will be particularly useful in this regard, as the thick disk and halo present a very different dynamical environment from globular clusters.

Additional abundance measurements of BHB stars will also prove useful, in order to better understand the details of the diffusion mechanism, the stability of BHB atmospheres, and the impact of diffusion on stellar flux distributions. If diffusion-driven metal enhancements do truly cause the $u$-band overluminosities, then the Str\"omgren survey of \cite{grundahl99}, which found a $u$-jump in all clusters surveyed, suggests that diffusion may be a ubiquitous feature of hotter BHB stars. Several HST STIS programs are underway to measure abundances of the hotter EHB stars, to see how far down the HB the metal enhancements extend. As the He\II\ convection zone predicted by \cite{sweigart02} approaches the stellar surface, it could reduce the effectiveness of the diffusion mechanisms, resulting in EHB stars with normal cluster metallicities. At yet higher $\Teff$, metal enhancements may reappear, as suggested by the photometric measurements of \cite{momany02}. Actual spectroscopic measurements of individual stars will be necessary to determine whether these scenarios are plausible.

%%%%%%%%%%%%%%%%%%%%%%%%%%%%%%%%%%%%%%%%

\clearpage
\acknowledgements

This paper covers the main results of my graduate thesis, submitted in partial fulfillment of the requirements for the degree of Doctor of Philosophy at the California Institute of Technology. I am deeply indebted to my thesis advisors, Judy Cohen and Jim McCarthy, for their guidance, patient tutelage, and generous allocations of Keck observing time towards this project. Thanks also go to thesis committee member George Djorgovski for suggesting BHB targets in M13 and contributing additional telescope time to this end. {\it Grazie} to Manuela Zoccali, Alejandra Recio-Blanco, Giampaolo Piotto, and Patrick Durrell for sharing spectroscopic and photometric results prior to publication, and a tip of the hat to Robert Kurucz, Michael Lemke, and Tim Pearson for making their computer codes readily available to the community. Clearly, this work would not have been possible without the prodigious collecting area of the Keck I telescope and the exquisite spectral resolution of the HIRES spectrograph. I applaud Jerry Nelson, Gerry Smith, Steve Vogt, and many others for creating such marvelous machines, and salute a bevy of Keck observing assistants, including Joel Aycock, Teresa Chelminiak, Barbara Schaefer, and Terry Stickel, for making them function properly for us. I acknowledge the W. M. Keck Foundation for funding the Keck Observatories, as well as specific financial support from NSF Grant AST-9819614, as well as support for HST GO proposal number 8617, which was provided by NASA through a grant from the Space Telescope Science Institute, which is operated by the Association of Universities for Research in Astronomy, incorporated, under NASA contract NAS5-26555. Lastly, I thank the anonymous referee for a thorough critique of the manuscripts, and for making many constructive suggestions that led to a much more comprehensive and complete paper.

%%%%%%%%%%%%%%%%%%%%%%%%%%%%%%%%%%%%%%%%

%%%%%%%%%%%%%%%%%%%%%%%%%%%%%%%%%%%%%%%%%%%%%%%%%%%%%%%%%%%%%%%%%

\newpage

\clearpage

%%%%%%%%%%%%%%%%%%%%%%%%%%%%%%%%%%%%%%%%%%%%%%%%%%%%%%%%%%%%%%%%%

\newpage

\begin{deluxetable}{lcccccc}
\tablecaption{Parameters for program clusters. \label{cluster-parameters}}
\tablewidth{0pt}
\tablehead{
cluster name		&[Fe/H]		&$E(\bv)$	&$\left( m-M \right)_0$		&$V_{\rm HB}$	&$v_r \,\, ({\rm km} \, {\rm s}^{-1})$		&$N_{\rm obs}$
}
\startdata
M3~~(NGC~5272)	&$-1.57$		&0.01	&15.12	&15.68	&$-148.5$	&\fn 6	\\
M13 (NGC~6205)	&$-1.54$		&0.02	&14.48	&15.05	&$-246.6$	&13		\\
M15 (NGC~7078)	&$-2.25$		&0.10	&15.37	&15.83	&$-107.3$	&18		\\
M68 (NGC~4590)	&$-2.06$		&0.05	&15.19	&15.68	&\fn$-94.3$	&11		\\
M92 (NGC~6341)	&$-2.29$		&0.02	&14.64	&15.10	&$-122.2$	&22		\\
NGC~288			&$-1.07$		&0.03	&14.69	&15.30	&\fn$-46.6$	&\fn 4	\\
\enddata
\end{deluxetable}

\begin{deluxetable}{lll}
\tablecaption{Finding charts for program stars. \label{finding-charts}}
\tablewidth{0pt}
\tablehead{
cluster	&star names	&reference
}
\startdata
M3		&B			&\cite{sandage53}, Figure 6					\\
M13		&IV			&\cite{arp55a}, Figure 2						\\
M13		&J11		&\cite{arp55b}, Plate 13					\\
M13		&SA			&\cite{savedoff56}, Figure 2					\\
M13		&WF			&Figure \ref{m13-wfpc} (this paper)			\\
M15		&B			&\cite{buonanno83a}, Figures 1--4				\\
M68		&W			&\cite{walker94}, Plate 18 \& Figure \ref{m68-fc} (this paper) 	\\
M92		&IV, VI, X, XII	&\cite{sandage66}, Figure 1					\\
M92		&B			&\cite{buonanno83b}, Figures 1--4				\\
NGC~288	&B			&\cite{buonanno84}, Figures 1--4				\\
\enddata
\end{deluxetable}

\begin{deluxetable}{ccccccc}
\tablecaption{HIRES grating configurations. \label{hires-config}}
\tabletypesize{\small}
\tablewidth{0pt}
\tablehead{
setting	&cross-disp.	&cross-disp.		&echelle			&echelle		&total $\lambda$	&continuous $\lambda$ 	\\[-2pt]
code		&grating 		&angle (deg)		&angle (deg)		&orders		&coverage (\AA)	&coverage (\AA) 	
}
\startdata
A		&RED		&$-0.202$		&$-0.400$		&54--82		&4304--6631 		&4304--5190 	\\
B		&RED		&$-0.492$		&$-0.121$		&57--91		&3885--6292 		&3885--5123	\\
C		&RED 		&$-0.429$		&$-0.037$		&56--87		&4066--6408		&4066--5126	\\
D		&UV			&\phm{-} 1.352	&\phm{-} 0.000	&65--88		&4020--5520		&4020--5126	\\
E		&UV			&\phm{-} 1.354	&\phm{-} 0.172	&66--90		&3936--5442		&3936--5059	\\
F		&UV			&\phm{-} 1.194	&\phm{-} 0.000	&67--91		&3888--5356		&3888--5126	\\
G		&RED		&$-0.816$		&\phm{-} 0.000	&61--91		&3888--5884		&3888--5127	\\
\enddata
\end{deluxetable}

\begin{deluxetable}{lccc}
\tablecaption{HIRES observations of BHB stars. \label{hires-obs}}
\tabletypesize{\footnotesize}
\tablewidth{0pt}
\tablehead{
\multicolumn{2}{l}{\sl civil date (observers)}		&slit \&		&mean	\\
\quad cluster/star		&exposures			&grating		&$S/N$	
}
\startdata
\multicolumn{2}{l}{\sl 1995 Oct 04--06 (Cohen, Shopbell)} \\
\quad M92/XII-9		&$3\times1200$ sec	&C5, A		&62		\\
\quad M92/XII-1		&$3\times1200$ sec	&C5, A		&72		\\
\quad M92/X-22		&$3\times1200$ sec	&C5, A		&65		\\
\quad M92/IV-27		&$3\times1200$ sec	&C5, A		&66		\\
\quad M92/IV-17		&$3\times1500$ sec	&C5, A		&42		\\[5pt]

\multicolumn{2}{l}{\sl 1997 Jan 07 (Behr, McCarthy)} \\
\quad M3/B518		&$3\times1200$ sec	&C5, B		&38		\\
\enddata
\end{deluxetable}

\begin{deluxetable}{lccc}
\tablecaption{Ratios of spectral resolution values derived from night sky emission lines (ns) versus arc line instrumental profiles (ip) . \label{nse-cmp}}
\tablewidth{0pt}
\tabletypesize{\scriptsize}
\tablehead{
$\lambda$~(\AA)	&$\langle R_{\rm ns}/R_{\rm ip} \rangle$	&rms$(R_{\rm ns}/R_{\rm ip})$		&$N_{\rm ns}$	
}
\startdata
5197.9		&1.05		&0.15		&15		\\
5564.0		&1.04		&0.11		&\fn 7	\\
5589.1		&1.05		&0.05		&\fn 5	\\
5915.3		&0.94		&0.05		&19		\\
5924.7		&0.98		&0.16		&13		\\
5932.9		&0.65		&0.03		&21		\\
5945.9		&1.27		&0.09		&\fn 6	\\
6170.6		&1.03		&0.09		&17		\\
6234.3		&0.96		&0.06		&13		\\
6257.9		&0.94		&0.06		&23		\\
6265.2		&0.77		&0.06		&22		\\
6287.4		&0.92		&0.05		&11		\\
6321.4		&0.99		&0.09		&14		\\
6348.5		&0.86		&0.19		&12		\\
6363.8		&0.84		&0.02		&11		\\
\enddata
\end{deluxetable}

\begin{deluxetable}{lccc}
\tablecaption{Predicted change in instrumental profile (i.p.) FWHM as a function of seeing FWHM due to slit ``underfilling''. \label{seeing-test}}
\tablewidth{0pt}
\tabletypesize{\scriptsize}
\tablehead{
				&				&seeing-corrected		&percentage	\\
i.p. FWHM 			&seeing FWHM 		&i.p. FWHM 			&change
}
\startdata
$6.46 \kms$		&0.7 arcsec		&$5.84 \kms$			&$-9.6$\%	\\
(C1 slit)			&0.8 arcsec		&$6.03 \kms$			&$-6.6$\%	\\
				&0.9 arcsec		&$6.18 \kms$			&$-4.3$\%	\\
				&1.0 arcsec		&$6.30 \kms$			&$-2.5$\%	\\
				&1.1 arcsec		&$6.38 \kms$			&$-1.2$\%	\\[4pt]
				
$8.56 \kms$		&0.7 arcsec		&$6.66 \kms$			&$-22.2$\%	\\
(C5 slit)			&0.8 arcsec		&$7.05 \kms$			&$-17.6$\%	\\
				&0.9 arcsec		&$7.36 \kms$			&$-13.9$\%	\\
				&1.0 arcsec		&$7.62 \kms$			&$-11.0$\%	\\
				&1.1 arcsec		&$7.82 \kms$			&\fn$-8.6$\%	\\
\enddata
\end{deluxetable}

\begin{deluxetable}{lcccccl}
\tablecaption{Photometry for BHB stars. \label{photometry}}
\tablewidth{0pt}
\tabletypesize{\scriptsize}
\tablehead{
star					&$V$		&\MV			&\BV				&\VI					&other				&ref.	
}
\startdata
M3/B125 (I-I-30)		&15.86		&0.74 \e 0.10		&\fmi 0.075 \e 0.02	&\fmi 0.049 \e 0.01	&\nodata				&1,6,9	\\
M3/B244 (I-II-66)		&15.96		&0.84 \e 0.10 		&\mi 0.01 \e 0.02		&\mi 0.062 \e 0.02		&\nodata				&1,9	\\
M3/B445 (I-I-56)		&16.34		&1.22 \e 0.10		&\mi 0.01 \e 0.02		&\mi 0.012 \e 0.01		&\nodata				&6,9	\\
M3/B518 (I-III-22)		&15.70		&0.58 \e 0.10		&\fmi 0.08 \e 0.02		&\nodata				&\nodata				&1		\\
M3/B831 (I-IV-78)		&15.88		&0.76 \e 0.10		&\fmi 0.00 \e 0.02		&\nodata				&\nodata				&1		\\
M3/B1241 (I-IV-18)		&15.99		&0.87 \e 0.10		&\mi 0.01 \e 0.02		&\nodata				&\nodata				&1		\\

M13/IV-83			&15.22		&0.74 \e 0.10		&\fmi 0.06 \e 0.02		&\nodata				&\nodata				&2		\\
M13/J11				&15.00		&0.52 \e 0.10		&\fmi 0.17 \e 0.02		&\nodata				&\nodata				&3		\\
\enddata
\tablecomments{references: 1. \cite{buonanno94}, 2. \cite{peterson95}, 3. \cite{sandage69}, 4. \cite{zoccali98},
5. \cite{buonanno85}, 6. \cite{rosenberg00a, rosenberg00b}, 7. \cite{walker94}, 8. \cite{buonanno84}, 9. \cite{ferraro97b}, 
10. \cite{cohen97}, 11. \cite{alcaino80}, 12. \cite{olszewski84}}
\end{deluxetable}

\begin{deluxetable}{llcccccccccccccccccccccc}
\rotate
\tablecaption{Atomic lines used for abundance analysis \label{linelist}}
\tabletypesize{\scriptsize}
\tablewidth{0pt}
\tablehead{
				&		&			&			&\multicolumn{3}{l}{star index:} \\
$\lambda$~(\AA)	&ion		&$\loggf$	&$\chi$~(eV)	&01 &02 &03 &04 &05 &06 &07 &08 &09 &10 &11 &12 &13 &14 &15 &16 &17 &18 &19 &20
}
\startdata
3888.650			&He\I  	&$-$0.714	&19.821 		&  	&  	&  	&  	&  	&  	&  	&  	&  	&  	&x 	&  	&  	&x 	&  	&x 	&  	&x 	&x  	\\
3889.132			&Mn\II 	&$-$2.385	&\phn 5.374 	&  	&  	&  	&  	&  	&  	&  	&  	&  	&  	&  	&  	&  	&  	&x 	&  	&  	&  	&   	\\
3892.293			&S\II 	&$-$0.060	&16.136 		&  	&  	&  	&  	&  	&  	&  	&  	&  	&  	&x 	&  	&  	&  	&  	&  	&  	&  	&   	\\
3894.073			&Co\I  	& 0.100		&\phn 1.049 	&  	&  	&  	&x 	&  	&  	&  	&  	&  	&  	&  	&  	&  	&  	&  	&  	&  	&  	&   	\\
3895.656			&Fe\I  	&$-$1.670	&\phn 0.110 	&  	&  	&  	&x 	&  	&  	&  	&  	&  	&  	&  	&  	&  	&  	&  	&  	&  	&  	&   	\\
3897.618			&Mn\II 	&$-$1.697	&\phn 5.381 	&  	&  	&  	&  	&  	&  	&  	&  	&  	&  	&  	&  	&  	&  	&x 	&  	&x 	&  	&   	\\
3898.067			&Mn\II 	&$-$2.060	&\phn 5.567 	&  	&  	&  	&  	&  	&  	&  	&  	&  	&  	&  	&  	&  	&  	&x 	&  	&  	&  	&   	\\
3899.707			&Fe\I  	&$-$1.531	&\phn 0.087 	&  	&  	&  	&x 	&  	&  	&  	&  	&  	&  	&  	&  	&  	&  	&  	&  	&  	&  	&   	\\
\enddata
\tablecomments{star indices: 1=M3/B125, 2=M3/B244, 3=M3/B445, 4=M3/B518, 5=M3/B831, 6=M3/B1241, 7=M13/IV-83, 8=M13/J11, 9=M13/SA113, 10=M13/SA404, 11=M13/WF2-2541, 12=M13/WF2-2692, 13=M13/WF2-3035, 14=M13/WF2-3123, 15=M13/WF2-820, 16=M13/WF3-548, 17=M13/WF3-1718, 18=M13/WF4-3085, 19=M13/WF4-3485, 20=M15/B78 \\
Full table, with all 74 stars and all 716 lines, available in electronic version.
}
\end{deluxetable}

\begin{deluxetable}{lllllll}
\tablecaption{Derived photospheric parameters for target stars. \label{phot-solutions}}
\rotate
\tabletypesize{\tiny}
\tablewidth{0pt}
\tablehead{
				&									&							&						&						&						&heliocentric \\
cluster/star		&$\Teff$, $\logg$ diagnostics				&\quad $\Teff $ (K)				&$\;\logg$ (cgs)			&$\xi$ diagnostics			&$\xi$ (km s${}^{-1}$)		&$v_r$ (km s${}^{-1}$)
}
\startdata
M3/B125			&$M_V$, \BV, \VI, Ti~II, Fe~II			&\fn 8937 \pl 169 \mi 102		&3.43 \pl 0.07 \mi 0.12 		&Mg~I, Ti~II, Fe~II			&1.87 \pl 0.97 \mi 0.95		 &\mi 149.3 \e 1.2		\\
M3/B244			&$M_V$, \BV							&\fn 9408 \pl 597 \mi 419		&3.48 \pl 0.09 \mi 0.09	 	&Fe~II					&0.75 \pl 1.57 \mi 0.75	 	&\mi 147.4 \e 4.2 		\\
M3/B445			&$M_V$, \BV, \VI						&10047 \pl 365 \mi 209			&3.71 \pl 0.07 \mi 0.09	 	&						&2.00 \pl 2.00 \mi 2.00	 	&\mi 155.1 \e 7.7 		\\
M3/B518			&$M_V$, \BV, Mg~I, Ti~II, Fe~I\&II		&\fn 8224 \pl 227 \mi 275		&3.21 \pl 0.09 \mi 0.09	 	&Ti~II, Fe~I\&II			&1.72 \pl 1.07 \mi 0.88	 	&\mi 149.1 \e 0.8 		\\
M3/B831			&$M_V$, \BV, Ti~II, Fe~I\&II				&\fn 9148 \pl 507 \mi 371		&3.41 \pl 0.09 \mi 0.08	 	&Mg~II, Ti~II, Fe~II			&1.57 \pl 0.82 \mi 0.86	 	&\mi 142.6 \e 0.8 		\\
M3/B1241		&$M_V$, \BV, Ti~II, Fe~II				&\fn 9600 \pl 582 \mi 644		&3.50 \pl 0.10 \mi 0.10		&Mg~I\&II, Fe~II			&1.06 \pl 1.27 \mi 1.06	 	&\mi 146.0 \e 1.4 		\\
M13/IV-83		&$M_V$, \BV, Ti~II, Fe~II				&\fn 8450 \pl 322 \mi 174		&3.28 \pl 0.11 \mi 0.07	 	&Mg~I, Ti~II, Fe~II			&2.99 \pl 0.93 \mi 0.68	 	&\mi 252.3 \e 1.7		\\
M13/J11			&$M_V$, \BV, Ca~I, Ti~II, Fe~I\&II			&\fn 7681 \pl 129 \mi 129		&3.08 \pl 0.07 \mi 0.08	 	&Mg~I, Ca~I, Ti~II, Fe~I\&II	&4.07 \pl 0.53 \mi 0.56	 	&\mi 246.4 \e 1.0		\\
\enddata
\end{deluxetable}

\begin{deluxetable}{lrrcccccccccccccc}
\rotate
\tablecaption{Error analysis of derived chemical abundance values \label{eps-results}}
\tabletypesize{\tiny}
\tablewidth{0pt}
\tablehead{
species	&$\logeps$	&[X/H]		&$N_{\rm lines}$	&$+\sigma_{\rm fit}$	&$-\sigma_{\rm fit}$	&$\sigma_{\rm Teff}$	&$-\sigma_{\rm Teff}$	
			&$+\sigma_{\rm logg}$	&$-\sigma_{\rm logg}$	&$+\sigma_\xi$		&$-\sigma_\xi$		&$+\sigma_{\rm vsini}$	&$-\sigma_{\rm vsini}$	&$\sigma_{\rm gf}$	
			&$+\sigma_{\rm total}$	&$-\sigma_{\rm total}$
}
\startdata
\sidehead{\sl M13/IV-83 :}
C\I		&$<$5.743	&$<$\mi 2.86	&1 	&\nodata		&\nodata		&\nodata		&\nodata		&\nodata		&\nodata		&\nodata		&\nodata		&\nodata		&\nodata		&\nodata	&\nodata		&\nodata			\\
N\I		&$<$6.598	&$<$\mi 1.40	&2 	&\nodata		&\nodata		&\nodata		&\nodata		&\nodata		&\nodata		&\nodata		&\nodata		&\nodata		&\nodata		&\nodata	&\nodata		&\nodata			\\
O\I		&$<$7.776	&$<$\mi 1.15	&2 	&\nodata		&\nodata		&\nodata		&\nodata		&\nodata		&\nodata		&\nodata		&\nodata		&\nodata		&\nodata		&\nodata	&\nodata		&\nodata			\\
Mg\I		&6.139		&\mi 1.44	&6 	&\pl 0.109	&\mi 0.111	&\pl 0.308	&\mi 0.147	&\mi 0.024	&\pl 0.013	&\mi 0.197	&\pl 0.191	&\pl 0.036	&\mi 0.036	&0.004	&\pl 0.380	&\mi 0.273		\\
Mg\II	&5.943		&\mi 1.64	&1 	&\pl 0.053	&\mi 0.053	&\pl 0.013	&\pl 0.013	&\pl 0.018	&\mi 0.010	&\mi 0.080	&\pl 0.104	&\pl 0.031	&\mi 0.025	&0.010	&\pl 0.124	&\mi 0.100		\\
Al\I		&4.115		&\mi 2.35	&1 	&\pl 0.251	&\mi 0.447	&\pl 0.339	&\mi 0.163	&\mi 0.017	&\pl 0.026	&\pl 0.007	&\pl 0.020	&\pl 0.046	&\mi 0.021	&0.013	&\pl 0.426	&\mi 0.477		\\
Al\II		&5.676		&\mi 0.79	&1 	&\pl 0.835	&\mi 0.480	&\pl 0.081	&\pl 0.156	&\pl 0.141	&\pl 0.081	&\pl 0.119	&\pl 0.131	&\pl 0.141	&\pl 0.106	&0.010	&\pl 0.904	&\mi 0.480		\\
Si\I		&$<$5.754	&$<$\mi 1.80	&2 	&\nodata		&\nodata		&\nodata		&\nodata		&\nodata		&\nodata		&\nodata		&\nodata		&\nodata		&\nodata		&\nodata	&\nodata		&\nodata			\\
\enddata
\tablecomments{Full table, with results from all 74 program stars, available in electronic version.}
\end{deluxetable}

\begin{deluxetable}{llllll}
\rotate
\tablecaption{Summary of chemical abundance results for program stars. \label{eps-summary}}
\tabletypesize{\scriptsize}
\tablewidth{0pt}
\tablehead{
cluster/star		&[He/H]						&[Mg/H]						&[P/H]						&[Ti/H]						&[Fe/H]
}
\startdata
M3/B125			&\mi 0.52 \pl 0.68 \mi 0.79 (2)	&\mi 1.45 \pl 0.14 \mi 0.14 (2) 	&\nodata						&\mi 1.23 \pl 0.18 \mi 0.16 (26) 	&\mi 1.51 \pl 0.20 \mi 0.18 (27) 	\\
M3/B244			&\pl 0.09 \pl 0.55 \mi 0.70 (5)	&\mi 1.51 \pl 0.15 \mi 0.26 (2) 	&\nodata						&\mi 1.43 \pl 0.41 \mi 0.43 (14) 	&\mi 1.44 \pl 0.29 \mi 0.42 (25) 	\\
M3/B445			&\mi 0.58 \pl 1.16 \mi 0.86 (3)	&\mi 1.57 \pl 0.27 \mi 0.29 (2) 	&\nodata						&\mi 1.74 \pl 0.87 \mi 0.47 (10) 	&\mi 2.25 \pl 0.47 \mi 0.72 (4) 	\\
M3/B518			&\nodata						&\mi 1.43 \pl 0.17 \mi 0.19 (2) 	&\nodata						&\mi 1.32 \pl 0.21 \mi 0.24 (46) 	&\mi 1.63 \pl 0.19 \mi 0.22 (31) 	\\
M3/B831			&\mi 0.92 \pl 0.75 \mi 0.81 (2)	&\mi 1.37 \pl 0.12 \mi 0.12 (2) 	&\nodata						&\mi 1.26 \pl 0.32 \mi 0.26 (23) 	&\mi 1.52 \pl 0.23 \mi 0.21 (24) 	\\
M3/B1241		&\mi 0.20 \pl 0.91 \mi 0.97 (10)	&\mi 1.29 \pl 0.17 \mi 0.23 (2) 	&\nodata						&\mi 1.15 \pl 0.38 \mi 0.46 (21) 	&\mi 1.38 \pl 0.25 \mi 0.36 (23) 	\\
M13/IV-83		&\nodata						&\mi 1.64 \pl 0.12 \mi 0.10 (1) 	&\nodata						&\mi 1.32 \pl 0.24 \mi 0.16 (55) 	&\mi 1.81 \pl 0.21 \mi 0.18 (46) 	\\
M13/J11			&\nodata						&\mi 1.61 \pl 0.09 \mi 0.09 (1) 	&\nodata						&\mi 1.40 \pl 0.10 \mi 0.10 (71) 	&\mi 1.81 \pl 0.11 \mi 0.11 (48) 	\\
\enddata
\end{deluxetable}

\begin{deluxetable}{lccccccccc}
\tablewidth{0pt}
\tablecaption{Rotation velocities and error estimates derived for program stars. \label{vsini-results}}
\tabletypesize{\scriptsize}
\rotate
\tablehead{
 	&$\vsini$	&$\sigma_{\rm fit}$	& 	&$\sigma_\xi$	&$\sigma_{\rm macro}$	&$\sigma_{\rm limb}$	&$\sigma_{\rm ip}$	&$\sigma_{\rm cont}$	&$\sigma_\Sigma{\rm sys}$	\\
star	&($\kms$)	&($\kms$)	&$N_{\rm lines}$	&($\kms$)	&($\kms$)	&($\kms$)	&($\kms$)	&($\kms$)	&($\kms$)}
\startdata
M3/B125    	&$14.72$    	&$+1.96$~$-1.86$    	&{\phn}72    	&$+0.14$~$-0.21$    	&$-0.06$~$+0.02$    	&$+0.07$~$-0.07$    	&$+1.13$~$-1.04$    	&$-0.10$~$+0.09$    	&$+1.45$~$-1.49$    	\\
M3/B244    	&$33.70$    	&$+6.04$~$-5.59$    	&{\phn}36    	&$+0.28$~$-0.09$    	&$-0.03$~$+0.01$    	&$+0.14$~$-0.18$    	&$+3.89$~$-2.98$    	&$-0.03$~$+0.07$    	&$+4.39$~$-3.31$    	\\
M3/B445    	&$31.97$    	&$+11.15$~$-8.51\fn$    	&{\phn}{\phn}7    	&$+0.42$~$-0.45$    	&$-0.02$~$+0.02$    	&$+0.17$~$-0.19$    	&$+0.90$~$-1.49$    	&$-0.06$~$+0.03$    	&$+1.53$~$-2.21$    	\\
M3/B518    	&${\phn}5.87$    	&$+2.08$~$-2.46$    	&136    	&$-0.44$~$-0.10$    	&$-0.25$~$+0.14$    	&$+0.03$~$-0.02$    	&$+0.74$~$-0.74$    	&$-0.67$~$+0.51$    	&$+1.43$~$-2.22$    	\\
M3/B831    	&${\phn}9.99$    	&$+1.47$~$-1.37$    	&{\phn}60    	&$+0.00$~$-0.11$    	&$-0.08$~$+0.03$    	&$+0.04$~$-0.06$    	&$+0.87$~$-0.76$    	&$-0.15$~$+0.12$    	&$+1.07$~$-1.16$    	\\
M3/B1241    	&${\phn}6.48$    	&$+3.43$~$-4.28$    	&{\phn}56    	&$-0.03$~$+0.00$    	&$-0.08$~$+0.05$    	&$+0.05$~$-0.01$    	&$+0.99$~$-0.92$    	&$-0.51$~$+0.39$    	&$+1.48$~$-1.56$    	\\
M13/IV-83    	&$32.85$    	&$+2.63$~$-2.46$    	&143    	&$+0.09$~$-0.34$    	&$-0.04$~$+0.03$    	&$+0.13$~$-0.16$    	&$+3.36$~$-2.62$    	&$-0.03$~$+0.05$    	&$+3.66$~$-3.20$    	\\
M13/J11    	&$22.23$    	&$+1.36$~$-1.32$    	&236    	&$+0.24$~$-0.56$    	&$-0.07$~$+0.06$    	&$+0.11$~$-0.12$    	&$+2.50$~$-0.14$    	&$-0.07$~$+0.05$    	&$+2.96$~$-0.95$    	\\
\enddata
\end{deluxetable}

\begin{deluxetable}{lcccccc}
\tablecaption{Kolmogorov-Smirnov tests of multimodal $\vrot$ distributions. \label{ks-distributions}}
\tabletypesize{\scriptsize}
\tablewidth{0pt}
\tablehead{
			&$\sigma_{\rm i}/v_{\rm i}$	&$v_1\:(f_1)$	&$v_2\:(f_2)$	&$v_3\:(f_3)$	&$P_{\rm KS}$
}
\startdata
\multicolumn{4}{l}{\sl M3 metal-poor (26 stars)} \\*
			&0.01	&32.9 (1.00)			&\nodata			&\nodata			&0.000		\\*[-2pt]
			&		&33.2 (0.30)			&12.2 (0.70)		&\nodata			&0.370		\\*[-2pt]
			&		&35.2 (0.01)			&19.0 (0.38)		&11.9 (0.62)		&0.582		\\*
			&0.03	&31.7 (1.00)			&\nodata			&\nodata			&0.000		\\*[-2pt]
			&		&31.9 (0.29)			&12.1 (0.71)		&\nodata			&0.458		\\*[-2pt]
			&		&34.0 (0.14)			&15.5 (0.50)		&10.4 (0.36)		&1.000		\\*
			&0.05	&30.6 (1.00)			&\nodata			&\nodata			&0.000		\\
\enddata
\end{deluxetable}

\begin{deluxetable}{llll}
\tablecaption{Radial velocities at multiple epochs \label{vr-compare}}
\tabletypesize{\scriptsize}
\tablewidth{0pt}
\tablehead{
star				&obs. date		&$v_r$ (km s${}^{-1}$)	&reference
}
\startdata
M13/IV-83		&1982/1983		&$-242.4 \pm 3.5$		&\cite{peterson83b}	\\[-2pt]
				&1998 Jun 27		&$-252.3 \pm 1.8$		&this paper			\\[2pt]
M13/J11			&1982/1983		&$-248.2 \pm 3.2$		&\cite{peterson83b}	\\[-2pt]
				&1998 Jun 27		&$-246.4 \pm 1.0$		&this paper			\\[2pt]
M15/B130		&1997 Aug 02		&$-114.6 \pm 0.6$		&this paper			\\[-2pt]
				&2000 Aug 01		&$-113.9 \pm 0.6$		&\cite{recioblanco02b}	\\[2pt]
M15/B218		&1997 Aug 03		&\fn$-95.1 \pm 2.5$	&this paper			\\[-2pt]
				&2000 Aug 01		&\fn$-94.6 \pm 0.5$	&\cite{recioblanco02b}	\\
\enddata
\end{deluxetable}

\end{document}